\documentclass[a4paper,twocolumn,11pt,accepted=2026-11-08
]{quantumarticle}
\pdfoutput=1
\usepackage[utf8]{inputenc}
\usepackage[english]{babel}
\usepackage[T1]{fontenc}
\usepackage{amsmath}
\usepackage{amsfonts}
\usepackage{siunitx}
\usepackage{upgreek}
\usepackage{amssymb}
\usepackage{booktabs}
\usepackage{float}
\usepackage{cite}
\usepackage{hyperref}
\usepackage[final]{changes}

\definechangesauthor[name={Authors}, color=blue]{A}

\begin{document}

\title{Detailed, interpretable characterization of mid-circuit measurement on a transmon qubit}

\author{Piper C. Wysocki}
\email{pcwysoc@sandia.gov}
\affiliation{Quantum Performance Laboratory, Sandia National Laboratories, Albuquerque, NM 87185 and Livermore, CA 94550, USA}
\affiliation{Department of Physics and Astronomy, University of New Mexico, Albuquerque, NM 87106, USA}

\author{Luke D. Burkhart}
\email{luke.burkhart@ll.mit.edu}
\affiliation{MIT Lincoln Laboratory, Lexington, MA 02421, USA}

\author{Madeline H. Morocco}
\affiliation{MIT Lincoln Laboratory, Lexington, MA 02421, USA}

\author{Corey I. Ostrove}
\affiliation{Quantum Performance Laboratory, Sandia National Laboratories, Albuquerque, NM 87185 and Livermore, CA 94550, USA}

\author{Riley J. Murray}
\affiliation{Quantum Performance Laboratory, Sandia National Laboratories, Albuquerque, NM 87185 and Livermore, CA 94550, USA}

\author{Tristan Brown}
\affiliation{MIT Lincoln Laboratory, Lexington, MA 02421, USA}

\author{Jeffrey M. Gertler}
\affiliation{MIT Lincoln Laboratory, Lexington, MA 02421, USA}

\author{David K. Kim}
\affiliation{MIT Lincoln Laboratory, Lexington, MA 02421, USA}

\author{Nathan E. Miller}
\affiliation{MIT Lincoln Laboratory, Lexington, MA 02421, USA}

\author{Bethany M. Niedzielski}
\affiliation{MIT Lincoln Laboratory, Lexington, MA 02421, USA}

\author{Katrina M. Sliwa}
\affiliation{MIT Lincoln Laboratory, Lexington, MA 02421, USA}

\author{Robin Blume-Kohout}
\affiliation{Quantum Performance Laboratory, Sandia National Laboratories, Albuquerque, NM 87185 and Livermore, CA 94550, USA}

\author{Gabriel O. Samach}
\affiliation{MIT Lincoln Laboratory, Lexington, MA 02421, USA}

\author{Mollie E. Schwartz}
\affiliation{MIT Lincoln Laboratory, Lexington, MA 02421, USA}

\author{Kenneth M. Rudinger}
\affiliation{Quantum Performance Laboratory, Sandia National Laboratories, Albuquerque, NM 87185 and Livermore, CA 94550, USA}

\begin{abstract}
Mid-circuit measurements (MCMs) are critical components of the quantum error correction protocols expected to enable utility-scale quantum computing. MCMs can be modeled by quantum instruments (a type of quantum operation or process), which can be characterized self-consistently using gate set tomography. However, experimentally estimated quantum instruments are often hard to interpret or relate to device physics. We address this challenge by adapting the error generator formalism---previously used to interpret noisy quantum gates by decomposing their error processes into physically meaningful sums of ``elementary errors''---to MCMs. We deploy our new analysis on a transmon qubit device to tease out and quantify error mechanisms including amplitude damping, readout error, and imperfect collapse. We examine in detail how the magnitudes of these errors vary with the readout pulse amplitude, recover the key features of dispersive readout predicted by theory, and show that these features can be modeled parsimoniously using a reduced model with just a few parameters.  
\end{abstract}

\begin{figure}[t!]
    \centering
    \includegraphics[width=1\linewidth]{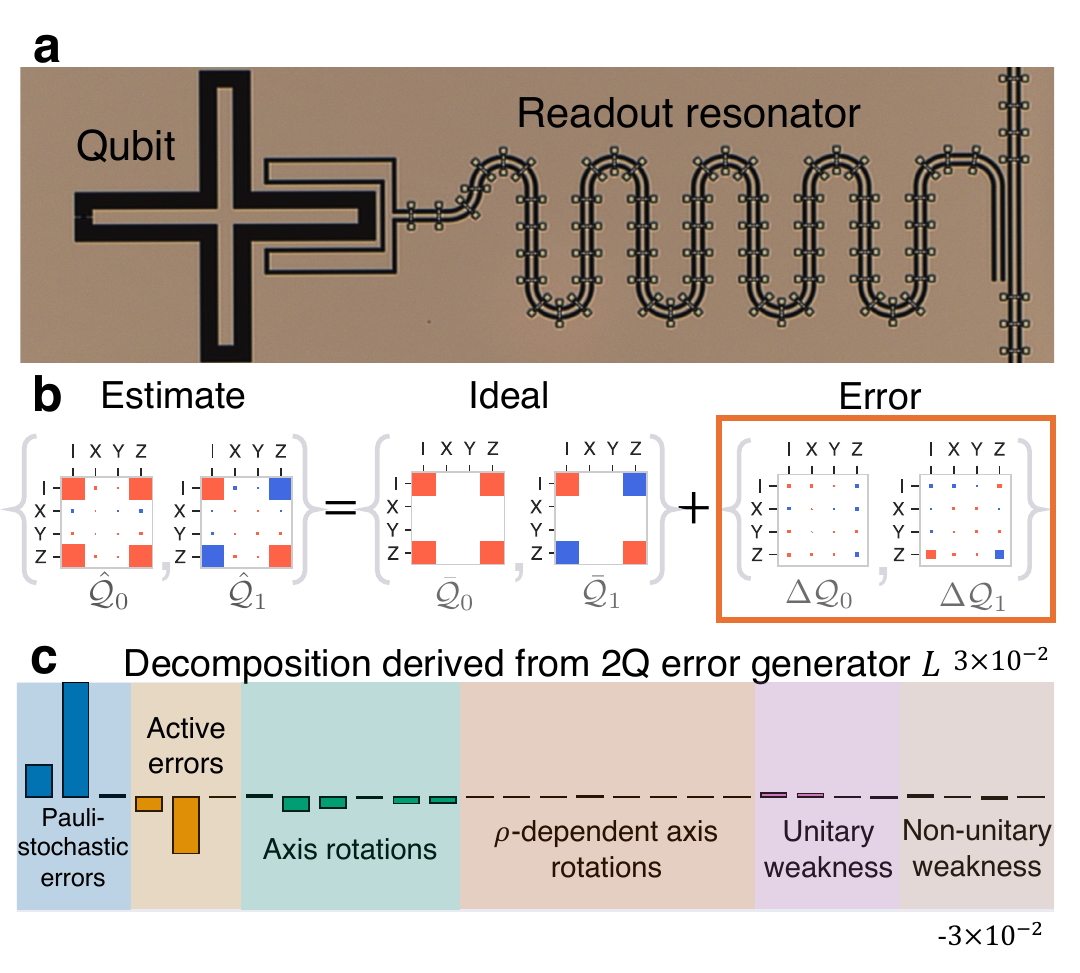}
    \caption{\textbf{Interpreting mid-circuit measurement (MCM) errors in experimental hardware.} (\textbf{a}) A noisy MCM implemented on a superconducting qubit device can be characterized using gate set tomography, which constructs an estimated quantum instrument $\hat{\mathcal{Q}}$. (\textbf{b}) We propose a new framework for interpreting $\hat{\mathcal{Q}}$. First, we write $\hat{\mathcal{Q}}=\bar{\mathcal{Q}}+\Delta \mathcal{Q}$, where $\bar{\mathcal{Q}}$ is the target (ideal) quantum instrument and $\Delta \mathcal{Q}$ is the deviation from $\bar{\mathcal{Q}}$ (error). (\textbf{c}) We leverage a circuit gadget representation of an MCM. Here, we treat MCM errors (red splat, left) as being produced by an error process $e^L$ (orange splat, right) acting on a joint physical and virtual qubit system. (\textbf{d}) Assuming a linearized regime, we use the elementary error generator representation \cite{blume-kohout2022a} of $L$ to establish an MCM error decomposition into physically interpretable error strengths. Notably, the error decomposition is sparse for most theory-derived MCM noise models as well as the experimental data shown here.} 
    \label{fig:summaryfig}
\end{figure}

High-fidelity \textit{mid-circuit measurements (MCMs)}, which read out individual qubits non-destructively during quantum computation, are critical enablers for quantum error correction (QEC) experiments on the path to fault-tolerant quantum computing \cite{krinner2022realizing, gupta2024probabilistic, bluvstein2024, acharya2024quantumerrorcorrectionsurface, harper2026characterising, besedin2026realizing}, and for quantum algorithms relying on dynamic circuits \cite{griffiths1996semiclassical,lu2022measurement, gupta2024probabilistic, baumer2024efficient}. Like all other logic operations, MCMs can fail, and thus constitute a source of error in quantum circuits that use them \cite{bluvstein2024, acharya2024quantumerrorcorrectionsurface,gupta2024probabilistic,harper2026characterising,Hothem2025measuring, Hines2025pauli,zhang2024generalizedcyclebenchmarkingalgorithm}. Quantifying and modeling MCM errors is essential for the development of utility-scale quantum computers. However, this task poses new challenges, as MCMs can display novel failure modes that impact circuit performance differently than gates or terminating measurements \cite{rudinger2022a}. 

MCM performance has previously been measured using variants of randomized benchmarking \cite{Hothem2025measuring} and Pauli noise learning \cite{Hines2025pauli,zhang2024generalizedcyclebenchmarkingalgorithm}.  These methods are lightweight and relatively easy to implement, but do not probe the nature of MCM errors in detail, and thus provide only limited debugging insight. Both methods use ``twirling'' to enforce a Pauli-stochastic error model: randomized benchmarking conglomerates all the resulting Pauli errors together into a single error metric for circuit layers that include MCMs, while Pauli noise learning estimates the rates of all the effective Pauli errors. However, these techniques have limitations and are not designed to probe the detailed physical behavior of MCMs.  They cannot separate the effects of, for example, readout errors that only affect the classical output bit, from $T_1$ decay that corrupts the post-measurement state. Indeed, in superconducting qubits, measurement calibration routinely involves a trade-off between these two error channels. The development of more detailed and flexible characterization techniques is needed to support debugging and mitigation of MCM error processes.

A noisy MCM can be fully modeled by an extension of the quantum operation (process) formalism known as a quantum instrument \cite{rudinger2022a, davies1970an}: a list of conditional quantum operations, written as transfer matrices and indexed by the possible outcomes of the MCM. The QI describing a particular MCM can be estimated using tomographic protocols \cite{wagner2020device, blumoff2016implementing, pereira2023parallel, stricker2022characterizing, rudinger2022a}. However, estimated quantum instruments can be difficult to interpret, which limits their diagnostic utility. Quantum logic gates, which are described by individual quantum operations/processes, have traditionally faced the same challenges, but the error generator framework \cite{blume-kohout2022a} eased the task of interpreting gate errors by enabling the decomposition of any logic gate's error process into a linear combination of physically meaningful elementary error generators, each with an associated error rate. This representation also facilitates construction of reduced models, which capture gate errors with fewer parameters and make it feasible to scale GST beyond two qubits. A similar approach has not yet been pursued for MCMs, leaving their error modes incompletely understood. 

In this paper, we adapt the error generator formalism to MCMs.  Doing so is nontrivial, because ideal MCM transfer matrices are non-invertible (unlike those for logic gates). We overcome this problem by constructing a perturbative representation of small Markovian errors in MCMs that closely resembles the elementary error generators of logic gates. These MCM error generators yield two benefits analogous to those provided by gate error generators. First, they enable interpretation of MCM errors by breaking them down into physically meaningful components. Second, they enable the creation of reduced models that simplify MCM characterization, paving the way for improved scalability in MCM characterization. We demonstrate both capabilities here, by analyzing and interpreting MCMs estimated using gate set tomography \cite{Nielsen2021gatesettomography,rudinger2022a} on a superconducting qubit device (Fig.~\ref{fig:summaryfig}).

The structure of this paper is as follows: Section~\ref{sec:math} introduces necessary mathematical preliminaries, including the quantum instrument formalism for modeling MCM errors. Additionally, it reviews the error generator formalism for gates, and shows that error generators do not straightforwardly extend to MCMs. Section~\ref{sec:MCM_error_gens} introduces our procedure for extracting error strengths from quantum instruments. Section~\ref{sec:experiment} describes the experimental setup and the GST experiments performed. In this section, we also analyze the experimental results to demonstrate the new framework's ability to interpret MCM errors and identify the significant error sources. Finally, Section~\ref{sec:discussion} discusses the impact and outlook of our results on MCM characterization.  

\section{Mathematical preliminaries}\label{sec:math}

\subsection{States, gates, and terminating measurements}
The state of a quantum computing register can be represented by a $d \times d$ density matrix $\rho$, where $d$ is the dimension of the quantum register's Hilbert space and $d=2^n$ if the register comprises $n$ qubits. The noisy implementation of a reversible quantum logic gate on the register can be represented by a \textit{quantum process} or completely positive, trace-preserving (CPTP) linear map on states, $G:\rho\rightarrow G[\rho]$. We represent quantum states as $d^2$-dimensional vectors $|\rho\rangle\!\rangle$ in the vector space of Hermitian $d\times d$ matrices, and CPTP maps as $d^2 \times d^2$ Pauli transfer matrices (PTMs) that act by left multiplication on $|\rho\rangle\!\rangle$, and whose matrix elements are given by
\begin{equation}
    [G]_{kl} = \frac{1}{d}\text{Tr}(\sigma_k G[\sigma_l])
\end{equation}
in terms of the $n$-qubit Pauli operators $\mathcal{P} = \{\sigma_1, \dots \sigma_{d^2}\}$. In this framework, terminating (destructive) measurements are represented by positive, operator-valued measures (POVMs) $\{E_i:\ i=1\ldots m\}$, where each $E_i\geq0$ is a $d\times d$ matrix called an \textit{effect}, which can be vectorized as a dual (row) vector $\langle\!\langle E_i|$, and $\sum_iE_i=I$. For convenience, we use $|\psi\rangle\!\rangle \equiv \big||\psi\rangle\!\langle \psi|\big\rangle\big\rangle$ as shorthand for the vectorized pure state projector $|\psi\rangle\!\langle \psi|$. In most cases in this paper, $|\psi\rangle\!\rangle = |0\rangle\!\rangle$ or $|1\rangle\!\rangle$.

These mathematical models of states, gates, and measurement outcomes can be used together to compute the outcome probability distribution for any experiment described by a quantum circuit.  For example, if a state $\rho$ is initialized and then gates $G_1$ and $G_2$ are performed in succession, the probability of observing measurement outcome $E_i$ is given by
\begin{equation}
    \mathrm{Pr}( E_i | \rho, G_1, G_2 ) = \langle\!\langle E_i| G_2 G_1 | \rho \rangle\!\rangle.
\end{equation}

\subsection{Quantum instruments (QIs)}

A noisy $m$-outcome \textit{mid-circuit} measurement can be modeled within this framework as a \textit{quantum instrument (QI)} \cite{hashim2025a, davies1970an, rudinger2022a}: a CPTP map from a quantum system to a combined quantum+classical system describing both the quantum register \textit{and} a classical $m$-state ``readout'' register. A QI can be described by a collection of $m$ completely positive, trace-non-increasing $d^2 \times d^2$ maps $\mathcal{Q} = \{\mathcal{Q}_0, \dots, \mathcal{Q}_{m-1}\}$, whose sum must be trace-preserving,
\begin{equation}
\text{Tr}\left(\sum_i^m \mathcal{Q}_i[\rho]\right) = \text{Tr}(\rho) \ \forall \ \rho.
\end{equation}
This condition ensures conservation of probability---the probabilities of the $m$ possible outcomes must sum to 1---because the probability of measuring outcome $i$ for an input state $\rho$ is given by 
\begin{equation}
    p_i = \text{Tr}(\mathcal{Q}_i[\rho]),
\end{equation}
and the output state $\rho'$ conditional on measuring outcome $i$ is 
\begin{equation} 
\rho_i' = \mathcal{Q}_i[\rho]/p_i.
\end{equation}
Each element $\mathcal{Q}_i$ of a QI can be written, just like the process describing a gate, as a $d^2 \times d^2$ PTM whose $(k,l)$-th matrix element is denoted as: 
\begin{equation}
    [\mathcal{Q}_i]_{kl} = \frac{1}{d}\text{Tr}(\sigma_k \mathcal{Q}_i[\sigma_l]). 
\end{equation}

\begin{figure}
    \centering
    \includegraphics[width=0.8\linewidth]{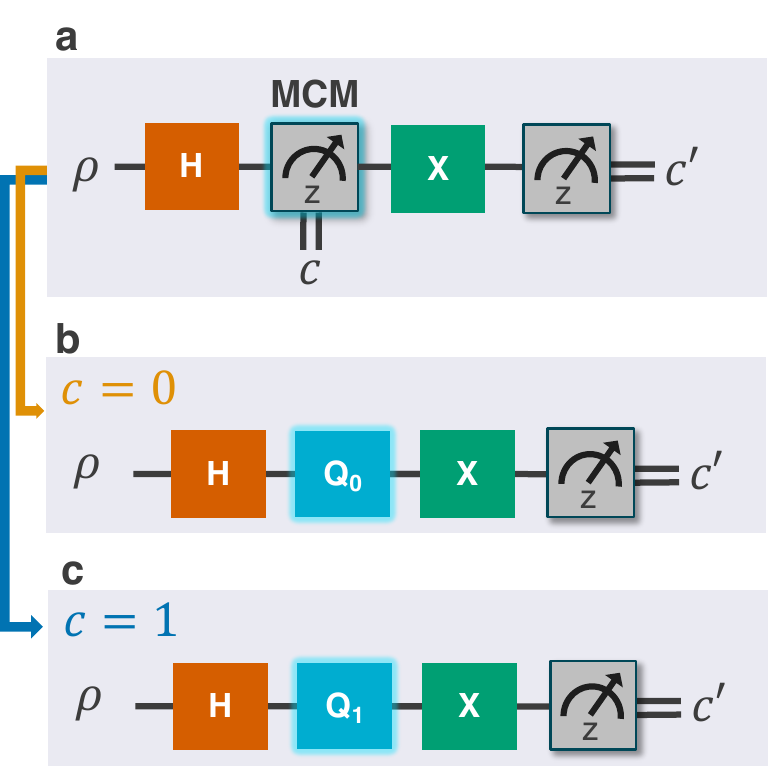}
    \caption{\textbf{Graphical interpretation of Eq.~\eqref{eq:QI_circuit_probs}.} (\textbf{a}) A quantum circuit containing an MCM can be interpreted as two different quantum processes, conditional on the measurement outcome: (\textbf{b}) $c=0$ and (\textbf{c}) $c=1$.}
    \label{fig:QI_explanation}
\end{figure}

For instance, Figure~\ref{fig:summaryfig}(b) shows the PTMs for the QI $\bar{\mathcal{Q}} = \{|0\rangle\!\rangle\! \langle\!\langle 0|, |1\rangle\!\rangle\! \langle\!\langle 1|\}$ that describes an ideal (i.e., noiseless) MCM in the computational basis. We can compute the probability of obtaining MCM outcome $i$ and terminating measurement outcome $j$ for an example circuit (Fig.~\ref{fig:QI_explanation}(a)) as follows: 

\begin{equation}
    p_{c=i, c'=j} = \langle\!\langle E_j |G_X \mathcal{Q}_i G_H |\rho \rangle\!\rangle
    \label{eq:QI_circuit_probs}
\end{equation} 
As illustrated in Fig.~\ref{fig:QI_explanation}(b-c), the outcome $c$ of the MCM whose probability we are calculating determines which element of the QI ($\mathcal{Q}_0$ or $\mathcal{Q}_1$) appears in Eq.~\eqref{eq:QI_circuit_probs}.

\subsection{Error generators}
PTMs can be difficult to interpret, making it challenging to identify specific error sources in quantum operations. The error generator framework enables extracting physically meaningful error rates from the estimated process describing a noisy gate \cite{blume-kohout2022a}. Representing a noisy gate in terms of \textit{elementary error generators (EEGs)} can provide insights into the physics of error mechanisms, enabling improvements in calibration and future fabrication efforts \cite{tanttu2024assessment, bartling2025universal, Stemp2024scalable, Madzik2022nuclear}. Here, an imperfect gate $G$ is modeled as being produced by a \textit{post-gate error process} $\mathcal{E}$ that follows an ideal target unitary $\bar{G}:$
\begin{equation}
    G = \mathcal{E}\bar{G}. 
\end{equation}
From $\mathcal{E}$, we can extract a \textit{post-gate error generator} $L = \log{[\mathcal{E}]} = \log[{G}\bar{G}^{-1}]$ that generates the error process $\mathcal{E}=e^L$ in the same sense that a Hamiltonian $H$ generates the unitary $U=e^{iH}$. This error generator can in turn be decomposed into a linear combination of EEGs, 

\begin{equation}
   L = \sum_i \epsilon_i L_i,
\end{equation}
where each $\epsilon_i$ is the \textit{rate} of a particular EEG $L_i$. The $2^n(2^n-1)$ distinct EEGs that can act on an $n$-qubit register can be classified into four sectors---\textit{Hamiltonian} ($H$), \textit{Pauli-stochastic} ($S$), \textit{Pauli-correlation} ($C$), and \textit{active} ($A$)---that transform an $n$-qubit density operator $\rho$ in distinct ways \cite{blume-kohout2022a}. Each EEG is indexed by one ($P$) or two ($P,Q$) Pauli operators, and their actions on a density matrix $\rho$ are given by \added{(also see Fig.~\ref{fig:taxonomy_original} in Appendix~\ref{appendix:equivalence_classes})}:
\begin{subequations}
\begin{equation}
H_P[\rho] = i(P\rho I - I \rho P)
\end{equation}
\begin{equation}
S_P[\rho] = P\rho P - I\rho I
\end{equation}
\begin{equation}
C_{P,Q}[\rho]  = P\rho Q + Q \rho P -\frac{1}{2}\{ \{P,Q\}, \rho \}
\end{equation}
\begin{equation}
A_{P,Q}[\rho] = i(P\rho Q - Q \rho P +\frac{1}{2}\{ [P,Q], \rho \}),
\end{equation}
\label{eq:eeg_types}
\end{subequations}
where $I$ indicates the identity operator. Hamiltonian ($H$) error generators produce coherent (unitary) error processes, which can often be attributed to imperfect calibration. Pauli-stochastic ($S$) generators produce depolarization, dephasing, and other kinds of random Pauli errors. Analyses and simulations of quantum error correcting codes usually analyze Pauli-stochastic error models \cite{knill2005quantum, chamberland2016thresholds,tuckett2020fault}; this ansatz can be enforced using randomized compiling \cite{wallman2016noise, hashim2023benchmarking}. Pauli-correlation ($C$) generators ``modify'' Pauli-stochastic errors to describe non-Pauli-stochastic processes, such as dephasing in the eigenbasis of $X+Y$. Active ($A$) generators produce a range of effects requiring feedback from the environment, such as affine shifts in the Bloch sphere.  The canonical example of such a process is amplitude damping or $T_1$ decay, produced by a linear combination of $S_X$, $S_Y$, and $A_{XY}$ generators.

The error generator formalism cannot be extended directly to MCMs because measurements are not reversible.  Ideal QIs $\bar{\mathcal{Q}}$ are generally not invertible---they annihilate observables that anticommute with the measured observable---and so it is not possible to define an error generator using $L=\log[{\mathcal{Q}\bar{\mathcal{Q}}^{-1}]}$.  However, in the following section we show how to construct a perturbative representation of errors in MCMs that reproduces most of the desirable properties of the error generator representation for noisy gates.  It enables extraction and classification of the error strengths shown in Fig.~\ref{fig:summaryfig}(d). These error strengths provide new insights into the mechanisms of MCM errors and their relative importance.

\section{Extending error generators to MCMs}\label{sec:MCM_error_gens}
\begin{figure}[t!]
    \centering
    \includegraphics[width=1\linewidth]{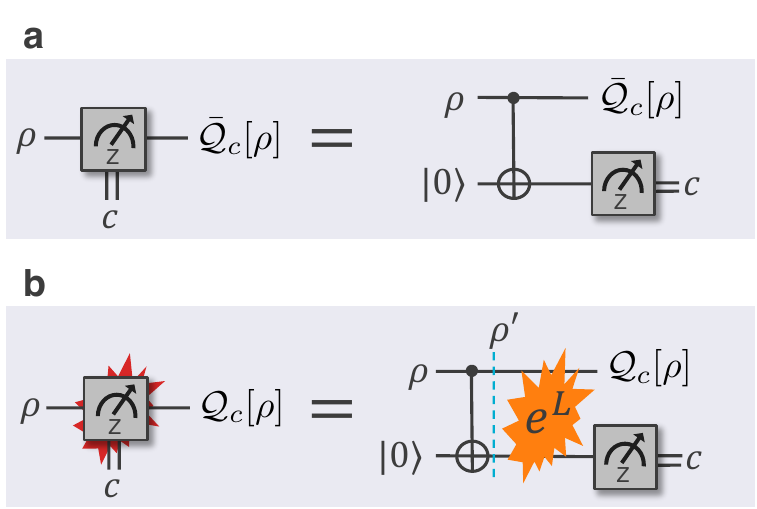}
    \caption{\textbf{A single-qubit MCM (left) can be written as a circuit gadget (right) on two qubits with no mid-circuit measurement}. (\textbf{a}) The noiseless case $\bar{\mathcal{Q}}_c[\rho]$: a virtual qubit is initialized ideally in $|0\rangle$ followed by a CNOT gate with the physical qubit as control and the virtual qubit as target. An ideal terminating measurement on the virtual qubit leaves the physical qubit in an output state conditional on the measurement outcome $c$. (\textbf{b}) The noisy case $\mathcal{Q}_c[\rho]$ (MCM with red splat): a post-gate error process generated by $L$ (orange splat) corrupts the CNOT. The dashed line indicates the noiseless initial state $\rho'$ of the physical and virtual auxiliary qubits and is given in Eq.~\eqref{jointstate}.}    \label{fig:auxiliary_pic}
\end{figure}

We begin with the observation that arbitrary errors in MCMs \textit{can} be represented using error generators---but for a larger system. Any single-qubit MCM can be modeled using the circuit gadget shown in Fig.~\ref{fig:auxiliary_pic}(a). This circuit comprises (1) error-free initialization of a virtual qubit in $|0\rangle$, (2) a CNOT operation from the physical qubit to the virtual qubit, and (3) an error-free terminating measurement of the virtual qubit. Now, arbitrary errors in the MCM can be modeled by an error process $\mathcal{E} = e^L$---where $L$ is a two-qubit error generator, depicted by the orange splat in Fig.~\ref{fig:auxiliary_pic}(b)---\textit{within} the gadget, immediately after the CNOT. 

An important subtlety is that the QI described by Fig.~\ref{fig:auxiliary_pic}(b) is not a fully faithful representation of the gadget error process $e^L$.  It cannot be, because a single-qubit QI can be described using just $2^2\times 2-2^2 =28$ real parameters, while a 2-qubit error process has $4^2\times (4^2-1)=240$ degrees of freedom. This mapping is therefore many-to-one---for each possible variation in the QI, there is an entire subspace of variations in $e^L$ that could produce it.  Because this mapping is linear, it can be captured by a two-qubit-error-process-to-one-qubit-instrument map $\mathcal{I}=\{\mathcal{I}_0, \mathcal{I}_1\}$. $\mathcal{I}$ ``crunches'' any gadget error process $e^L$ on the physical+virtual qubit system down to a QI acting on the physical qubit. The action of each element of $\mathcal{I}$ is defined as 

\begin{equation}
\begin{split}
    \mathcal{I}_c(e^L) &\equiv (I \otimes  \langle\!\langle c|) \cdot e^L  \cdot \text{CNOT}\cdot (I\otimes |0\rangle\!\rangle) \\ 
    &= \mathcal{Q}_c, 
\label{eq:QIfromError}
\end{split}
\end{equation}
where the single-qubit map $\mathcal{Q}_c$ is implicitly a function of the two-qubit error generator $L$. To illustrate how Eq.~\eqref{eq:QIfromError} works, we consider two example errors in MCMs and construct QIs for them from two-qubit EEGs. \begin{figure*}[t!]
    \centering
    \includegraphics[width=1\linewidth]{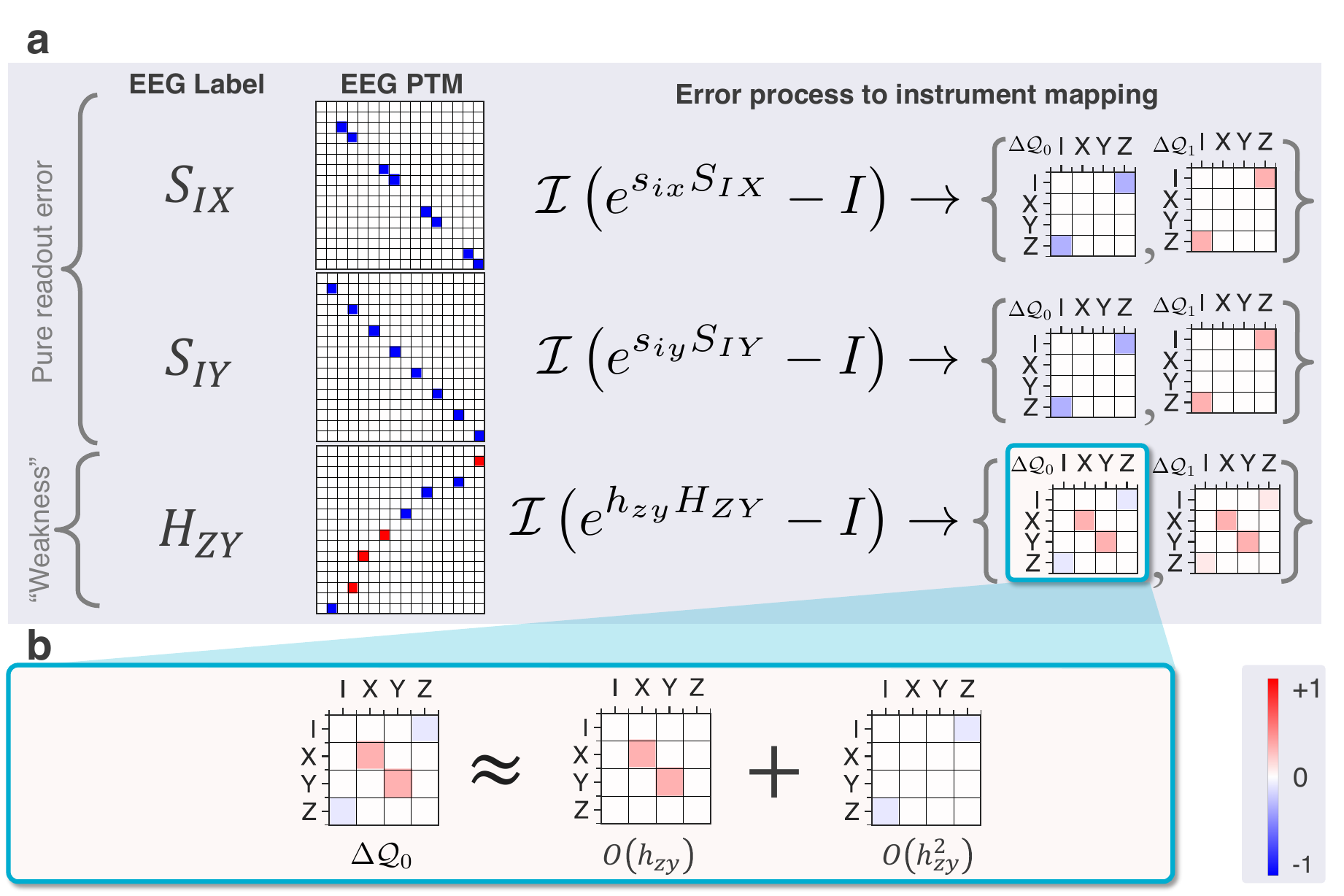}
\caption{\textbf{Two-qubit error processes are mapped to single-qubit instrument errors [Eq.~\eqref{eq:QIfromError}].} (\textbf{a}) The deviation $\Delta \mathcal{Q} = \mathcal{Q}-\bar{\mathcal{Q}}$ between the ideal QI $\bar{\mathcal{Q}}$ produced by an identity ``error'' process and the noisy QIs $\mathcal{Q}$ generated by elementary error generators (EEGs) $S_{IX}$ with weight $s_{ix}$, $S_{IY}$ with weight $s_{iy}$, and $H_{zy}$ with weight $h_{zy}$. The two-qubit PTM has Pauli labels $II$,$IX$, $\dots$$ZZ$. Note that $\Delta Q^{(s_{ix})} = \Delta \mathcal{Q}^{(s_{iy})}\ne \Delta \mathcal{Q}^{(h_{zy})}$. (\textbf{b}) $H_{ZY}$ produces both first-order effects (inner diagonal sub-block) and second-order effects (anti-diagonal corners), with strengths $\mathcal{O}(h_{zy})$ and $\mathcal{O}(h_{zy}^2)$ respectively. Further, the second-order effects produce the same deviation as the first-order effects of $S_{IX}$ and $S_{IY}$ in (\textbf{a}).}
    \label{fig:errorgen_to_QI}
\end{figure*}

\begin{itemize}
\item \textbf{Pure readout error}:  an input state of $|0\rangle$ is incorrectly read out as $c=1$, but the post-MCM quantum state remains $|0\rangle$ (and/or vice versa). This error process often arises from incorrect classification of an analog readout signal due to finite signal-to-noise ratio (as we later see in Fig.~\ref{fig:error_mechanisms}(b)). In the gadget picture, pure readout errors can, for example, be produced by a Pauli X error on the virtual qubit after the CNOT application and prior to measurement; this Pauli X error is generated by the Pauli-stochastic EEG $S_{IX}$.
\item \textbf{``Weakness''}: the quantum state of the register is not completely collapsed into an eigenstate of the measured observable, such that anticommuting observables are not completely annihilated \cite{clerk2010, hatridge2013}. Incomplete collapse occurs, in the gadget picture, if the CNOT gate does not properly rotate the virtual qubit conditional on the state of the physical qubit.  This can be produced by a two-qubit unitary following the CNOT, and can be generated, for example, by the Hamiltonian EEG $H_{ZY}$.
\end{itemize}
As we noted above, each error can be generated in multiple ways. For example, pure readout error can equally well be produced by a Pauli $Y$ error on the virtual qubit, which is generated by the EEG $S_{IY}$. Since the physical qubit and virtual qubit are in an eigenstate of $ZZ$ following the noiseless CNOT, the error processes generated by Pauli $X$ and Pauli $Y$ errors are indistinguishable once mapped to the single-qubit QI. Figure~\ref{fig:errorgen_to_QI} provides a schematic representation of the mapping between an error process and a QI, showing the two-qubit PTM, the resulting QI, and the deviation $\Delta\mathcal{Q}$ for $S_{IX}$, $S_{IY}$, and $H_{ZY}$ EEGs. We can see that although $S_{IX}$ and $S_{IY}$ correspond to different error processes in the two-qubit gadget picture, they produce the same deviation $\Delta \mathcal{Q}$ in the single-qubit QI.

This many-to-one mapping reflects a gauge freedom in the representation of QI errors by 2-qubit error generators.  To avoid confusion with the well-known (and more pernicious) gauge freedom present in gate sets \cite{Nielsen2021gatesettomography}, we refer to this new gauge as the \textit{MCM gauge}. Physically distinct deviations in the QI correspond to \textit{equivalence classes} (subspaces) of the 240-dimensional space of two-qubit error processes.  In the regime of \textit{small} errors, where $e^L \approx I + L$, this space is isomorphic to the 240-dimensional space spanned by the 2-qubit EEGs, and the MCM gauge-equivalence classes can be constructed explicitly. We provide this construction in Appendix~\ref{appendix:equivalence_classes}, and then in Appendix~\ref{appendix:classification} we suggest a classification of these MCM deviations that is inspired by the classification of gate error generators and facilitates their connection to device physics models.

Critically, our equivalence class formalism only provides quantitative insights when second-order (and higher-order) effects can reasonably be expected to be small. For example, Figure~\ref{fig:errorgen_to_QI}(b) illustrates that the Hamiltonian generator $H_{ZY}$ can produce the same error process at second order as stochastic generators $S_{IX}$ and $S_{IY}$ at first order. Second-order effects are negligible when the total MCM error is small, which parallels the assumption made in Ref. \cite{blume-kohout2022a} for gate errors. As we will see in Section~\ref{sec:error_decomp}, extracted error strengths can still yield qualitative insights in larger error regimes, particularly when only a handful of extracted error strengths are large and higher-order effects can easily be predicted. In the next section, we discuss the practical application of these quantities to understanding the performance of an MCM in a transmon qubit.

\section{Experimental demonstration}\label{sec:experiment}

\begin{figure*}[t!]
    \centering
    \includegraphics[width=1\linewidth]{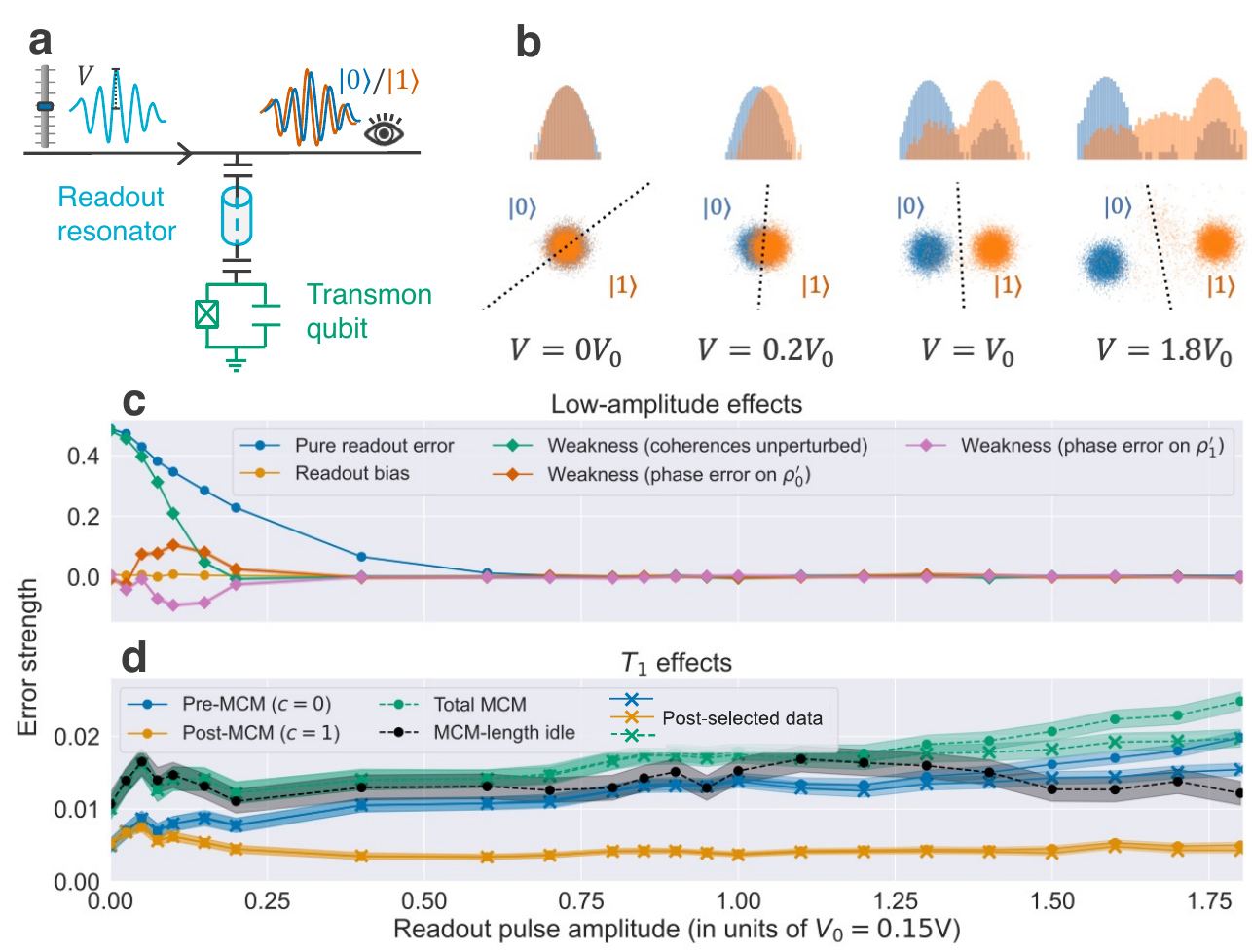}
    \caption{\textbf{MCM error mechanisms as a function of the readout pulse's amplitude $V$.} (\textbf{a}) A mid-circuit measurement is implemented through dispersive readout. This setup enables non-destructive state measurement of individual qubits by leveraging the interaction between the qubit (green) and a readout resonator (blue). The readout amplitude is systematically swept to control the drive strength, allowing for the investigation of measurement errors as a function of amplitude. (\textbf{b}) We plot single-shot IQ data from an experiment used to calibrate the MCM for four representative values of $V$ as point clouds (bottom) and log-scaled histograms of counts projected on the I axis (top). (\textbf{c})-(\textbf{d}) To better understand the error mechanisms present in the MCM, we extract error strengths from the GST estimates as a function of $V$ using the error decomposition. Shaded regions show $2\sigma$ error bars, which in some cases are smaller than the plotted line width. (\textbf{c}) Errors that only appear in the low-amplitude regime, where the signal-to-noise ratio is low: weakness (blue, orange, and pink lines) and pure readout error/measurement bias (yellow and blue lines). Panel (\textbf{d}) Errors that can be attributed to $T_1$ effects, occurring either before or after the MCM. The sum of all $T_1$ effects (dashed green line) can be compared with the $T_1$ effects predicted by idling for the same period as the measurement (dashed black line). In the absence of added $T_1$ effects from the MCM, these two quantities should be equal. ``X''-shaped markers show data where leakage events have been removed via post-selection; see Section~\ref{ssec:leakage}.}
    \label{fig:error_mechanisms}
\end{figure*}

We investigate the dynamics of an MCM in a transmon qubit, where the mid-circuit and terminating measurements are implemented by dispersive readout using a capacitively-coupled transmission line resonator [Ref. \cite{Koch2007charge}, Fig.~\ref{fig:error_mechanisms}(a)]. We place the readout probe tone halfway between the resonator frequencies when the qubit is induced by qubit states $|0\rangle$ ($\omega_0$) and $|1\rangle$ ($\omega_1$) so that the resonator response is symmetric for both qubit states. The qubit-state-dependent frequency shift of the resonator is $\chi_\mathrm{01} = \omega_1 - \omega_0 = -2\pi \times 0.26  $ MHz and the resonator linewidth is $\kappa  = 2\pi \times 0.16$ MHz. The qubit excited state lifetime $T_1$ is approximately $90$ $\upmu$s.

The device measured in this work was intentionally designed with weak readout coupling to enable study of qubit coherence due to materials quality. The parameters $\chi_\mathrm{01}$ and $\kappa$ are small compared to those of devices designed for state-of-the-art measurement fidelity. Our experiments demonstrate that the techniques of Section~\ref{sec:MCM_error_gens} can be applied broadly not just to optimized devices. However, this device's achievable measurement rate is relatively slow. We reduced the measurement time by using a readout probe pulse shape similar to the one presented in Ref. \cite{McClure2016rapid}, which drives the resonator more strongly at the start and end of the pulse, enhancing the photon number ring-up and ring-down. The measurement pulse is $2.3$ $\upmu$s in duration, followed by an additional 2 $\upmu$s delay after the pulse to ensure the resonator has returned to its ground state. For more experimental details, see Appendix~\ref{experimental_details}.
 
Using \textit{gate set tomography} (GST) \cite{Nielsen2021gatesettomography,rudinger2022a}, we characterize how the MCM's behavior depends on its strength by varying the amplitude of the readout probe pulse. We define a “nominal” measurement operation, chosen heuristically to achieve minimal assignment error, which we use for all terminal measurements. The nominal measurement is defined by the peak voltage amplitude of its readout pulse as measured at the output of the waveform generator at room temperature, which is $V_0$ = $0.15$ volts. We sweep the peak voltage $V$ of the MCM pulse from $V=0$ (no readout pulse) to $V=1.8 \times V_0$ (where we expect to over-drive the readout resonator, introducing additional error mechanisms) to study the behavior of the dispersive readout. For each value of $V$ we perform a standard reference experiment preparing $|0\rangle$ and $|1\rangle$, and train a state assignment classifier independently for use in the corresponding GST experiment. The results of a subset of these reference experiments are shown in Fig.~\ref{fig:error_mechanisms}(b).

GST fully and self-consistently characterizes a gate set $\mathcal{G}$. Here, the gate set $\mathcal{G}$ includes state preparation in $|0\rangle$, terminating measurement in the computational basis, single-qubit gates $G_X$ and $G_Y$ corresponding to $\pi/2$ pulses around the x and y-axes, respectively, and an idle operation $G_I$ for the full duration of the MCM (which includes the post-MCM delay time). Additionally, the gate set includes the MCM itself. Each experiment consists of only 128 circuits (36 of which contain the MCM). This circuit list is identical to the one given in Ref. \cite{rudinger2022a} where GST with MCMs was first demonstrated. We take $N=8 \times 10^{3}$ shots of each circuit to obtain high precision estimates. We employ maximum likelihood estimation (MLE), implemented using \texttt{pyGSTi} \cite{osti_code-28250, Nielsen2020}, to find a best-fit model of the gate set, including the MCM.  MLE involves numerically maximizing the \textit{likelihood} function, defined as
\begin{equation} \label{eq:L}
    \mathcal{L} = \mathrm{Pr}(\mathrm{data}|\mathrm{model}),
\end{equation} 
by varying all the parameters of the gate set model.  The baseline CPTP model has 71 parameters---12 matrix elements for each of the 3 gates, 7 more for the state preparation and terminating measurement operations, and 28 for the MCM---but a 12-dimensional gauge freedom \cite{Nielsen2020} leaves only 59 degrees of freedom that influence the likelihood. 

\subsection{Error decomposition}\label{sec:error_decomp}
For each chosen value of $V$, we conduct a GST experiment as described above and perform an error decomposition on the estimated QI to extract the error strengths associated with each equivalence class of error generators. Each such class corresponds to a physically meaningful error process (Appendix~\ref{appendix:classification}). We can connect the observed features in the error decomposition to the underlying device physics visible in the raw readout data.

The per-measurement IQ voltage data in Fig.~\ref{fig:error_mechanisms}(b) illustrates that the separation of distributions corresponding to $|0\rangle$ and $|1\rangle$ is minimal at low amplitude, leading to poor distinguishability. As expected, the separation increases with amplitude, corresponding to an increase in state distinguishability and assignment fidelity. The extracted error strengths elucidate the effects of indistinguishability at low amplitude. In Fig.~\ref{fig:error_mechanisms}(c), we observe significant pure readout error (blue line) for $V < 0.6\times V_0$  and weakness (green line) for $V < 0.2\times V_0$ due to reduced measurement strength when the readout resonator is populated with fewer photons \cite{Gambetta2006qubit}. In addition to incomplete collapse of coherence, we also observe a deterministic phase shift on both the $\rho'_0$  and $\rho'_1$ output states, which results from the deterministic AC Stark shift of the dispersive Hamiltonian. 

At values of $V$ above $0.7 \times V_0$, the IQ distributions become increasingly distinguishable and we observe a corresponding reduction in the reported weakness and pure readout errors. In this regime, the dominant error mechanism is amplitude damping from $|1\rangle$ to $|0\rangle$ due to $T_1$ events. We distinguish two different amplitude damping processes, both shown in Fig.~\ref{fig:error_mechanisms}(d): one corresponding to an effective pre-MCM amplitude damping process (blue line) and the other to an effective post-MCM amplitude damping process (yellow line).

While both error processes correspond to a state transition from $|1\rangle$ to $|0\rangle$, pre-MCM T1 errors also manifest as an output bit $c=0$, 
while post-MCM errors lead to an output bit $c = 1$.
Because the measurement and assignment is a complex time-evolution process, the resultant error rates will be a nontrivial function of the dynamics of the readout resonator under stochastic discrete state transitions. These two error modes have distinctly different implications for QEC. Post-MCM $T_1$ effects can be mitigated by performing an unconditional reset operation following the MCM, while pre-MCM $T_1$ effects can introduce measurement errors during syndrome extraction. The corresponding effect is visible in the histograms in Fig.~\ref{fig:error_mechanisms}(b) at $V = V_0$, where the counts measured when preparing $|1\rangle$ deviate from a sum of Gaussians, with more counts at intermediate voltages. This is a common property of transmon readout that often limits measurement fidelity \cite{heinsoo2018}.

As the readout amplitude increases, we observe a gradual change in the strength of some $T_1$ decay error processes. We observe an increase in pre-MCM $T_1$ decay, while the post-MCM $T_1$ decay during the delay remains largely constant. Since the experiment includes an idle operation with the same duration as the measurement, we can compare the $T_1$ decay during idling (dashed black line) with the total observed $T_1$ decay during the MCM (dashed green line). If the MCM does not introduce additional $T_1$ events, these two quantities should be equal. Instead, they diverge at high $V$. This suggests that the $T_1$ decay rate increases with increasing $V$, as has been observed in other experiments \cite{thorbeck2024}.

However, the raw data indicates additional large-amplitude error effects. At the highest amplitudes we observe a significant, resolved third cluster in between the two corresponding to  $|0\rangle$ and  $|1\rangle$. The location of this cluster is consistent with leakage events corresponding to excitation of the transmon out of the computational manifold, which has been observed to occur as the number of photons populating the readout resonator increases \cite{thorbeck2024,dumas2024,sank2016}. 
\added{While in this experiment we do not implement control of the transmon's higher-energy excited state and cannot directly confirm that this cluster is unambiguously due to leakage, In Appendix~\ref{appendix:leakage_postselection} we present additional experimental evidence that the behavior is consistent with long-lived excitations out of the $\{|0\rangle, |1\rangle\}$ subspace. Thus we refer to it as "leakage," and in section 3.2 we analyze the impacts of this effect on the Markovian model.}
This third cluster is most prominent when the $|1\rangle$ state is prepared, indicating preferential leakage from $|1\rangle$ compared to $|0\rangle$. The orientation of the signal in the IQ plane also rotates slightly with amplitude, most likely due to the amplitude-dependent frequency shift of the resonator resulting from the Kerr nonlinearity, which does not impact the behavior of the QI when using the correctly trained classifier.

\subsection{Probing leakage at high readout amplitude}\label{ssec:leakage}
Since our MCM model assumes a two-level quantum system, it cannot account for errors that cause leakage out of the computational subspace, such as measurement-induced state transitions \cite{sank2016}. Consequently, out-of-model effects can lead to incorrect estimates of in-model error rates, because maximum likelihood estimation seeks to find the model parameters that best describe the data. To investigate how leakage may distort estimate of MCM error strengths, we perform the error decomposition on a subset of data where we have removed a significant fraction of what we suspect to be leakage events.

Leakage to higher levels of the transmon is expected to have a long lifetime, comparable to the $T_1$ time. Therefore, most leakage events which occur during the MCM will be detected by the terminating measurement. This observation allows us to post-select out the leaked shots without relying on the MCM IQ data. We use the terminating measurement IQ data from the dataset with the largest observed leakage cluster to train a three-state classifier, with outcomes ``0'', ``1'', and ``Leaked.'' We label this third cluster ``Leaked'' instead of ``2'' because our experiments cannot determine to which higher energy level of the transmon the leakage occurs. Since the amplitude of the terminating measurement is kept fixed, the post-selection process is independent of the MCM amplitude, and we can use this same classifier for all datasets.

For each value of $V$, we remove experimental shots in which the classifier assigned ``Leaked'' in the terminating measurement. This post-selection process removes between 1.4\% and 2.3\% of shots from any GST experiment, depending on the amplitude, but between 1.0\% and 8.5\% of shots from any individual circuit. This indicates that the probability of leakage is circuit-dependent, which is consistent with the observation that in Fig.~\ref{fig:error_mechanisms}(b), leakage appears to preferentially originate from the $|1\rangle$ state. See Appendix~\ref{appendix:leakage_postselection} for additional details on the post-selection procedure.

The error decomposition of post-selected GST results generally does not differ from that of the full GST results, with a key exception being the $T_1$ effects (`X' markers, Fig.~\ref{fig:error_mechanisms}(d)). We observe that the post-selected results exhibit a smaller rise in pre-MCM $T_1$ decay errors than the full results. This confirms that leakage errors are incorrectly attributed in part to amplitude damping in the full GST results.

This is likely due to the fact that the leakage is observed to occur more frequently from the $|1\rangle$ state, and the leakage in the full dataset is assigned a ``0'' outcome with roughly 50\% probability, so in some circuits, the effects of leakage \added{(e.g., $|1\rangle\rightarrow |\mathrm{leaked}\rangle\rightarrow \mathrm{measure} \ 0$)} and $|1\rangle \rightarrow |0\rangle$ decay are indistinguishable. \replaced{I}{For example, i}n a circuit that prepares $|1\rangle$ followed by an MCM and a terminal measurement, the probability of a ``0'' terminating measurement outcomes is \added{thus} proportional to both the amplitude damping and leakage probabilities.

As is often the case for out-of-model errors, the precise attribution of the errors by the model can depend on fine details, and care must be taken when interpreting such data. Nonetheless, even this simple modification of the analysis protocol allows disambiguation of the leakage and decay rates. It is likely that expanded models can be developed to better quantify this kind of incoherent leakage process; this will be the subject of future investigation. For the remainder of this paper, we elect not to perform post-selection and assess how well different models fit the complete datasets. 

\begin{figure*}[t]
    \centering
    \includegraphics[width=1\linewidth]{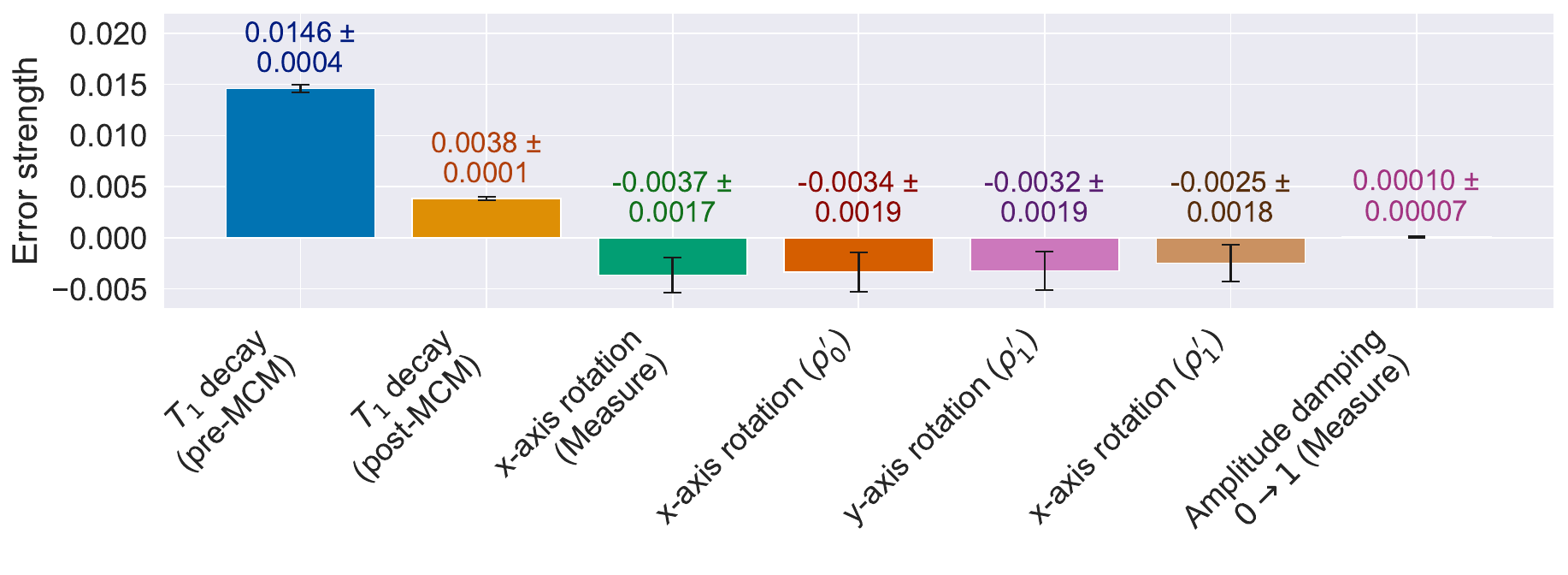}
    \caption{\textbf{Attributed MCM errors for the many-shot dataset.} 
    Error strengths of the 7 elementary deviations that are significant for the high-precision dataset at $V = V_0$ with $4 \times 10^4$ shots. For axis rotations, error strengths can be negative, as they correspond to the angle of the rotation. Only errors which are outside $2\sigma$ error bars are shown.}
    \label{fig:error_decomposition}  
\end{figure*}

\subsection{Assessing, enhancing, and simplifying the estimated model}\label{sec:goodness_of_fit}
Figure~\ref{fig:error_mechanisms} indicates that at least some error mechanisms, such as weakness, are strongly suppressed for sufficiently large readout amplitude. This indicates that errors in our device's MCM may be captured by a \textit{reduced model} describing simpler physics and fewer distinct errors than the full 28-parameter QI. Our analysis of the effects of leakage indicate that, at the nominal readout amplitude $V_0$, the leakage rate is small enough to not corrupt the estimate of the strength of $T_1$ processes. This absence of out-of-model leakage effects suggests that this is a viable setting at which to probe the potential of reduced models. We choose this operating point to perform an additional experiment taking $5\times$ as much data ($N=4 \times 10^{4}$ shots) and fit it to the CPTP model. Increasing the number of shots reduces the uncertainty in the error strength estimates and allows better detection of out-of-model effects. Appendix~\ref{app:gst_results} shows the full PTM estimates of this experiment. 

In Fig.~\ref{fig:error_decomposition}, we summarize the significant error strengths in this experiment. We observe that $T_1$ events continue to be dominant. Their distribution between pre- and post-MCM errors also remains comparable to the dataset with fewer shots. However, the higher precision of the estimates allows us to conclusively identify the presence of two other, much weaker error mechanisms: (1) amplitude damping from $|0\rangle$ to $|1\rangle$ caused by thermal relaxation at finite temperature and (2) pre- and post-MCM rotations of unknown origin. This highlights that the error landscape is more complex than commonly understood, and these second-order effects may become increasingly important to understand as MCM fidelities improve. 

Overall, in this larger dataset only 7 of the possible 28 parameters are significantly different from zero. In this section, we deploy statistical model selection to find a reduced model that still captures the relevant errors. In the process of doing so, we find that even the full QI model fails to fully explain the data---but just one additional parameter describing a non-Markovian (context-dependent) Stark shift enables the model to fit the data well.

\subsubsection{Model validation and non-Markovianity}\label{sec:non_markovianity}
We consider several models for the noisy MCM. To evaluate them, we fit each model to the data by finding the values of its parameters that maximize the likelihood $\mathcal{L}$ [Eq.~\eqref{eq:L}]. The maximized value $\mathcal{L}_{\mathrm{max}}(\mathrm{model})$ of the likelihood for each model can be used to measure the \textit{relative} fit quality of different models. Our baseline CPTP model is \textit{Markovian}: it describes each logic operation (initialization, terminating measurement, reversible gates, MCM) by a fixed-but-otherwise-general CPTP map that does not depend on the operation's context (e.g., what operation preceded it).  

Before evaluating reduced models constructed by eliminating parameters from the CPTP model, we evaluate whether the CPTP model itself is ``good''.  We do this by comparing its likelihood to that of a saturated or ``maximal'' model \cite{Nielsen2021gatesettomography} that assigns independent probabilities to every circuit outcome.  If our model is fully consistent with the data, then as $N_{\mathrm{shots}}\to\infty$ the loglikelihood ratio 
\begin{equation}
2\Delta \log{\mathcal{L}} \equiv 2\log\left[\frac{\mathcal{L}_{\mathrm{max}}(\mathrm{saturated})}{\mathcal{L}_{\mathrm{max}}(\mathrm{model})}\right]
\end{equation}
is a $\chi^2_k$ random variable \cite{wilks1938large}, where $k$ is the number of additional parameters in the saturated model compared to our model.  We can therefore quantify the observed \textit{model violation} \cite{Nielsen2021gatesettomography} by the number of standard deviations (of the $\chi^2_k$ distribution) by which the observed $\Delta \log \mathcal{L}$ exceeds its expected null value of $k$:
\begin{equation}
    N_{\sigma} \equiv \frac{2\Delta \log{\mathcal{L}}-k}{\sqrt{2k}}.
    \label{eq:n_sigma}
\end{equation}
Whenever $N_\sigma \gg 1$, data lies significantly outside the predictions of the model and we consider the model to be incorrect. In the context of the CPTP model, this indicates \textit{non-Markovian} behavior, since by construction the CPTP model can describe any Markovian dynamics.

\begin{figure*}[!t]
    \centering
    \includegraphics[width=1\linewidth]{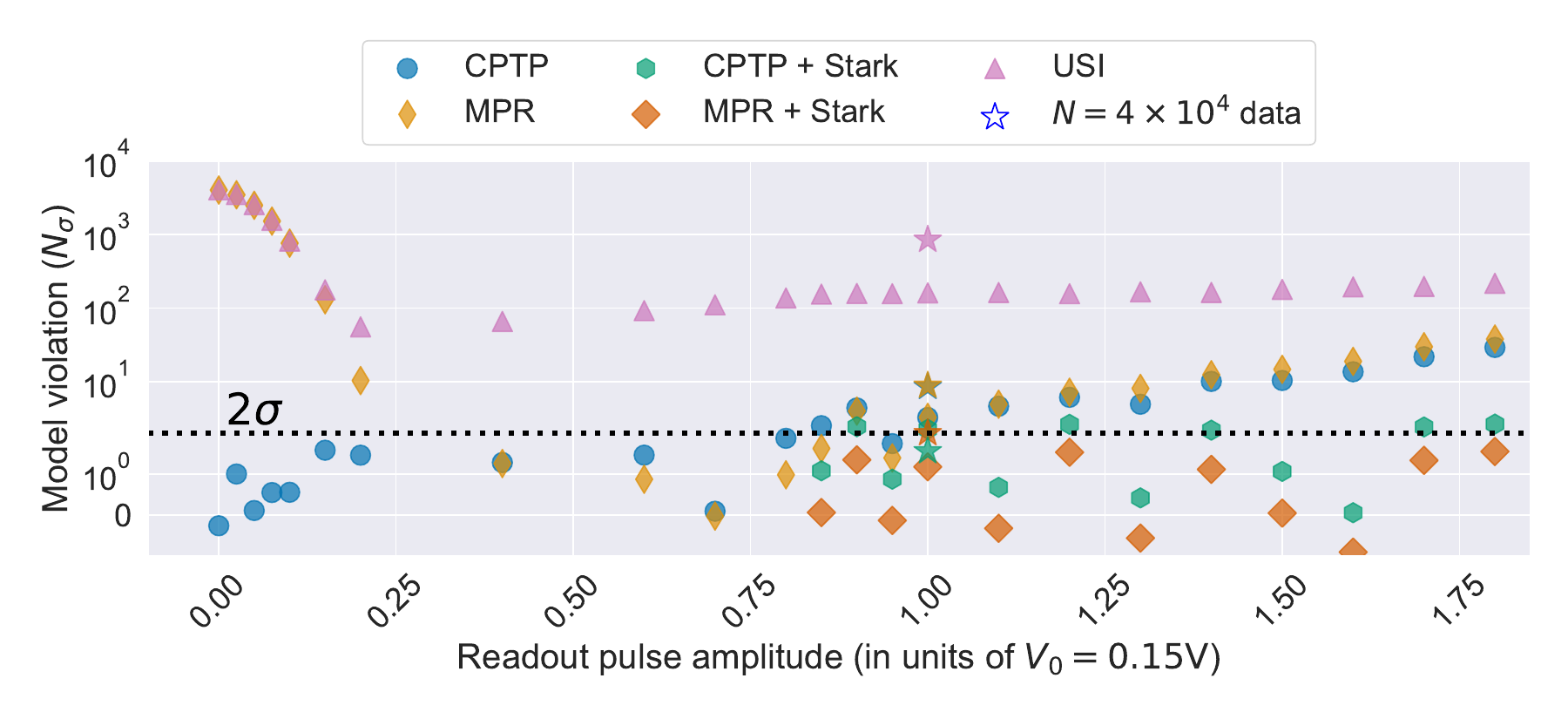}
    \caption{\textbf{Goodness-of-fit for MCM models.} We consider four models for the MCM: a full Markovian parameterization as a CPTP-constrained QI (``CPTP''), a uniform stochastic instrument (``USI''), and a measure-and-prepare instrument augmented with pure readout error (``MPR'').  We examine two further models (``CPTP + Stark'' and ``MPR + Stark'') where an additional parameter is added to capture the coherent Z phase error from the AC Stark shift on post-MCM gates. The first Stark shift model parametrizes the MCM as a CPTP-constrained QI and the second as an augmented measure-and-prepared instrument. For each of the five models, evidence for unmodeled error is quantified by number of standard deviations ($N_\sigma$) of model violation. A negative value for $N_\sigma$ indicates a value that many standard deviations \textit{below} the mean of the associated $\chi^2$ distribution for the null hypothesis. Stars indicate models fit to the dataset with $N=4 \times 10^{4}$ shots. Additional shots increase sensitivity to unmodeled error, so the increased $N_\sigma$ does not necessarily indicate more unmodeled error in this dataset. For each model, the increase in reported model violation from $N=8 \times 10^{3}$ shots to $N=4 \times 10^{4}$ shots is consistent with the increase in $N$.}
    \label{fig:goodness-of-fit}
\end{figure*}

To search for non-Markovian dynamics, we performed experiments and evaluated $N_\sigma$ for the CPTP model at many values of $V$ (blue circles in Fig.~\ref{fig:goodness-of-fit}). Two regimes emerge, separated by $V = 0.8\times V_0$. When $V < 0.8 \times V_0$, $N_\sigma ^{\mathrm{(CPTP)}} < 2$, indicating consistency with the Markovian CPTP model.\footnote{To be more precise, our experiments failed to reject the Markovian null hypothesis at the 95\% confidence level.  This is generally interpreted as ``no significant evidence for non-Markovianity''. This should not be taken as proof of Markovianity!  Our experience is that if enough data are taken, \textit{every} quantum computing experiment will display some non-Markovianity.} For most values of $V > 0.8\times V_0$, we observe $N_\sigma ^{\mathrm{(CPTP)}}> 5$, which is conclusive evidence of non-Markovian error. Inspection of the data (not shown) showed that only circuits containing MCMs were inconsistent with the CPTP model, which strongly suggests that the observed non-Markovianity is associated specifically with the MCM. To validate this hypothesis, we fit a gates-only model 
to a subset of the data excluding $36$ circuits containing MCMs. In all cases, $N^{\mathrm{(Gates)}}_\sigma < 3$, indicating little to no statistical evidence for non-Markovianity in the gates. We conclude that the MCM causes non-Markovian effects that grow with $V$ for $V>0.8\times V_0$.

Since non-Markovian behavior is clearly present for $V > 0.8\times V_0$, we attempt to model it, by augmenting the existing model with an additional parameter. We conjecture that the observed non-Markovianity is caused by photons that remain in the readout resonator even after the 2 $\mu$s post-MCM delay. There is a dispersive interaction between the readout resonator and the transmon, so these photons induce an AC Stark shift on the transmon, which generates a coherent Z phase error that is proportional to the average number of photons in the resonator \cite{Gambetta2006qubit}. In our experiment, these effects manifest as additional errors on the gates applied after the MCM \cite{rudinger2022a}, which is a form of non-Markovianity known as \textit{serial context dependence} (the action of a gate depends on preceding operations). The occupation of the readout resonator scales as $V^2$, so we expect this error to grow rapidly with $V$. 

In Appendix~\ref{appendix:stark}, we show how to augment the CPTP model so that it can capture this non-Markovian phase error, by adding just one additional parameter. We call this the ``CPTP+Stark'' model.  We fit the CPTP+Stark model to all the datasets where $N_\sigma^{\mathrm{CPTP}} > 2$. The results (green squares in Fig.~\ref{fig:goodness-of-fit}) show that adding the AC Stark effect reduces model violation significantly, but does not generally eliminate it. For large $V$, $N_\sigma$ is reduced by a factor of two to three.  We conclude that the CPTP+Stark model fits and explains the data much better than the CPTP model alone (see also Table~\ref{tab:fits}).  At $V=1.0\times V_0$, the CPTP+Stark model explains even the $N=4\times 10^4$ dataset well ($N_\sigma^{\mathrm{(CPTP+Stark)}}<2$).

\subsubsection{Simplifying the error model}\label{sec:reduced_models}
We now consider how to \textit{simplify} the model by removing extraneous parameters. Reduced models for gates have already been demonstrated \cite{tanttu2024assessment, Stemp2024scalable, Madzik2022nuclear}. We therefore focus on constructing reduced models for the MCM. The extracted error strengths (Fig.~\ref{fig:error_mechanisms}) indicate that many of the 28 deviations are statistically insignificant. We \textit{could} construct reduced models by iteratively setting small error strengths to zero, but this approach is uncomfortably close to data-mining and may not provide any \textit{physical} intuition regarding the MCM error processes in the experiment. Instead, we construct two candidate reduced models for MCM errors from first principles and evaluate their quality using the data. Ideally, a reduced model should capture the system's dynamics accurately and provide insight into the sources of error. Some reduced models also facilitate simulation of circuits containing MCMs---such as syndrome extraction circuits used in quantum error correction---which rapidly become computationally intractable unless simplifying approximations about MCM noise are made.

The first reduced model we consider is a \textit{uniform stochastic instrument} (USI), which generalizes the widely-used stochastic error model in quantum error correction to MCMs \cite{mclaren2023stochasticerrorsquantuminstruments, hines2024paulinoiselearningmidcircuit}. A single-qubit USI is mathematically represented as \cite{hines2024paulinoiselearningmidcircuit, beale2023randomized}: 

\begin{equation}
    \mathcal{Q}_c = \sum_{a,b\in\mathbb{Z}_2}q_{a,b}|c \oplus b\rangle\!\rangle \langle\!\langle c \oplus a|,
\end{equation}
where $|c\rangle\!\rangle \equiv ||c\rangle\!\langle c|\rangle\!\rangle$ and $\sum_{a,b\in\mathbb{Z}_2}q_{a,b}=1$. As such, three parameters are necessary to parameterize a USI, corresponding to the three possible bit-flip error patterns. 
\begin{table*}[t!]
    \centering
    \begin{tabular}{p{2.2cm}p{3cm}p{2cm}p{3.8cm}p{4cm}}
        \toprule
        Model Name & Total Parameters  &$2\Delta\log\mathcal{L}$& Model Violation $(N_{\sigma})$ & $\gamma(\text{Model, CPTP+Stark})$ \\ 
        \midrule
        CPTP+Stark & 60  & 166& 1.6& -- \\
        MPR+Stark & 43  &192& 2.0& 1.5\\
        MPR & 42  &313& 8.8 & 8.4\\
        CPTP & 59  &284& 8.5& 118\\
        USI & 34  &15290 & 830& 582\\ 
        \bottomrule
    \end{tabular}
    \caption{\textbf{How well five different MCM models fit the $N=4\times 10^4$ dataset.} Five models were fit to the same dataset.  A model's fit quality relative to a baseline model is determined by (1) its number of free parameters, and (2) its loglikelihood ratio \added{$2\Delta \log{\mathcal{L}}$} relative to the baseline model.  Here, we evaluate each model relative to a saturated (maximal) model using $N_\sigma$ \added{[Eq.~\eqref{eq:n_sigma}]}, and we evaluate the 4 \textit{reduced} models \added{defined in Sec.~\ref{sec:reduced_models}} relative to CPTP+Stark \deleted{(which contains all of the reduced models)} using the evidence ratio \replaced{$\gamma(\text{Model}, \text{CPTP+Stark})$}{$\gamma$} \added{[Eq.~\eqref{eq:evidence_ratio}]}. \added{CPTP+Stark denotes the CPTP model augmented by a single Stark-shift parameter in Sec.~\ref{sec:non_markovianity}.} The Akaike information criterion (AIC) suggests a reduced model yields more accurate predictions iff $\gamma<2$, indicating that MPR+Stark is the ``best'' model of MCM errors for this dataset.}
    \label{tab:fits}
\end{table*}
The second reduced model we consider is a \textit{measure-and-prepare} instrument, extended to include pure readout error. We denote this extended model as MPR in Fig.~\ref{fig:goodness-of-fit}. Standard measure-and-prepare instruments describe any MCM where no quantum information is transmitted through the measurement process, i.e., a process where the output state is given by $\mathcal{Q}_c\left[|0\rangle\!\rangle\right]$ \cite{zhang2024generalizedcyclebenchmarkingalgorithm, rudinger2022a}, instead of the input-state-dependent $\mathcal{Q}_c\left[|\rho\rangle\!\rangle\right]$. By definition, such QIs destroy all entanglement between the qubit and other qubits in the device \cite{Horodecki2003entanglement}. For a single-qubit MCM, this parameterization has $10$ non-gauge parameters, and can be implemented using two error processes $\mathcal{A}$ and $\mathcal{B}$ acting, respectively, on a terminating measurement and a conditional re-initialization: 
\begin{equation}
    \mathcal{Q}_c = \mathcal{A}|c\rangle\!\rangle \langle\!\langle c|\mathcal{B}.
\end{equation}
As it stands, this model cannot capture pure readout error, which is a small, but not insignificant error source in all datasets due to the finite signal-to-noise ratio. We therefore define ``MPR'' as an extended measure-and-prepare model with 1 additional parameter ($p$, the probability of reporting the incorrect outcome $c \oplus 1$) to capture pure readout error,
\begin{equation}
    \mathcal{Q}_c = \mathcal{A}\left((1-p)|c\rangle\!\rangle \langle\!\langle c|+p|c \oplus1\rangle\!\rangle \langle\!\langle c\oplus 1|\right)\mathcal{B}.
\end{equation}
We also define an augmented version of this model, MPR+Stark, that includes the one additional parameter introduced in the previous subsection to capture non-Markovianity induced by AC Stark shifts.  We use this model in place of the MPR model for the values of $V$ where $N^{\mathrm{(CPTP)}}_\sigma > 2$. In Fig.~\ref{fig:goodness-of-fit}, the wide orange diamonds correspond to the case where this additional Stark shift parameter must be added to the model, while the narrow orange diamonds represent cases where no additional parameter is required.  

We quantify the validity and utility of reduced models using:
\begin{enumerate}
    \item the $N_\sigma$ statistic introduced in the previous subsection, which quantifies our statistical confidence that the model is incorrect, and
    \item the \textit{evidence ratio} $\gamma$ \cite{Nielsen2021efficient,rudinger2021experimental}, which quantifies expected predictive accuracy.
\end{enumerate}  
The evidence ratio between two models $\mathcal{M}_A$ and $\mathcal{M}_B$, where reduced model $\mathcal{M}_B$ has fewer parameters than $\mathcal{M}_A$, is defined as
\begin{equation}
    \gamma(\mathcal{M}_A, \mathcal{M}_B) = \frac{2(\log{\mathcal{L_B}-\log{\mathcal{L_A})}}}{k_A-k_B},
    \label{eq:evidence_ratio}
\end{equation}
where $\mathcal{L}_i$ and $k_i$ are (respectively) the maximized likelihood of the $i$th model and the number of parameters in the $i$th model. A convenient interpretation of the evidence ratio is given by the Akaike information criterion (AIC) \cite{akaike1974a}, which states that a smaller model $\mathcal{M}_B$ has greater predictive accuracy iff $\gamma(\mathcal{M}_A, \mathcal{M}_B) < 2$. In other words, each additional parameter in the larger model is justified only if its inclusion reduces the loglikelihood by $\geq 2$ units.

Fitting a USI model to data (lilac triangles, Fig.~\ref{fig:goodness-of-fit}) yields very high $N^{\mathrm{(USI)}}_\sigma > 50$ for all values of $V$. This is not unexpected, because USIs are by construction highly symmetrized, and do not include any error mechanism that would be eliminated by randomized compiling \cite{wallman2016noise}, as is performed in Refs. \cite{beale2023randomized, hashim2023benchmarking}. Amplitude damping, such as in $T_1$ relaxation, is a simple example of an error process which cannot be captured by a USI because it is not uniform, and acts asymmetrically on $|0\rangle$ and $|1\rangle.$  Our results confirm that USI models should not be relied on to predict experiments unless all MCMs have been subject to randomized compiling.

There is only one regime for $V$ where the extended measure-and-prepare model fails to be as predictive as the full Markovian model, when quantified by either the evidence ratios $\gamma(\mathcal{M}_{\mathrm{CPTP}}, M_{\mathrm{MPR}})$ or $\gamma(\mathcal{M}_{\mathrm{CPTP+Stark}}, \mathcal{M}_{\mathrm{MPR+Stark}})$ as appropriate: when we expect non-trivial weakness ($V \le 0.2\times V_0$). The additional parameters in the full Markovian model are essential for fitting weakness. In fact, using such a model we are able to exactly capture the observed dynamics at $V = V_0$ in the $N=4 \times 10^4$ dataset, without any model violation ($N_\sigma^{\mathrm{(MPR + Stark)}} < 2$ and $\gamma(\mathcal{M}_{\mathrm{CPTP+Stark}}, \mathcal{M}_{\mathrm{MPR+Stark}})<2$ in Table~\ref{tab:fits}). Ultimately, for most values of $V$, we have captured the system dynamics with seventeen fewer parameters needed to model the MCM compared to the full QI, and only one additional parameter is required to capture most of the non-Markovian error.  

\section{Discussion}\label{sec:discussion}
The advancement of high-fidelity mid-circuit measurements is crucial for progress towards utility-scale quantum computation. Previously, GST has been shown to be a powerful tool for understanding the performance of mid-circuit measurements by reconstructing a full QI description of a mid-circuit measurement \cite{rudinger2022a}. For a single-qubit gate set that includes a mid-circuit measurement, GST typically requires only around $100$ circuits and often provides a simpler and more economical alternative for extracting critical quantities than conducting numerous distinct experiments. Furthermore, GST ensures that estimates are self-consistent and allows for statistically robust methods for model verification.

Our contribution provides a procedure for decomposing estimated QIs into \textit{interpretable} error strengths, enabling tomographic estimates of mid-circuit measurements (including those made via GST) to be more effectively used to diagnose and address physical error modes. We demonstrate that we can capture the main features of dispersive readout—including amplitude damping, non-collapse, and pure readout error —without making any assumptions beyond the Markovianity of the underlying noise model.

Moreover, our work is platform-agnostic, offering a new framework for interpreting QIs across various quantum computing architectures, beyond just superconducting qubits. This versatility opens avenues for future research, including the exploration of novel readout schemes \cite{kerman2013quantum,didier2015fast,billangeon2015circuit,wang2025longitudinal, jones2025sensor}. Additionally, our methodology can be extended to study other error correction primitives, such as parity checks \cite{wysocki2025full}; this promises to enhance our understanding of error dynamics in fault-tolerant hardware. In these cases, the auxiliary qubit(s) may not be virtual, and the observable error mechanisms in the QI can realistically be driven by multiple underlying error processes.

We further demonstrate how we can improve the versatility of GST of mid-circuit measurements by combining reduced models inspired by estimated error strengths with selective augmentation to capture the AC Stark shift, a non-Markovian error source. This compact model can be used to more efficiently simulate mid-circuit measurements without sacrificing predictive accuracy. In addition to informing our understanding of the performance of logical qubits under different mid-circuit measurement error models, such reduced models could potentially enable detailed characterization of large syndrome extraction circuits.

\section*{Data availability}
\added{The experimental data presented in this paper is available upon reasonable request from the corresponding authors.}

\section*{Author Contributions}
\added{P.C.W., L.D.B., G.O.S., and K.M.R. were the primary technical contributors to this work and contributed throughout the project to conceptual development, analysis and interpretation, and preparation of the manuscript. M.H.M. led the photon number analysis. R.B.-K. provided the original seed idea for using the CNOT gadget and provided substantial feedback on the manuscript. C.I.O. and K.M.S. provided technical, code, and experimental advice and input through project discussions. R.J.M. developed code for the quantum instruments used in the GST analysis. T.B., J.M.G., and N.E.M. contributed to experimental infrastructure. D.K.K. and B.M.N. fabricated the specific device used in this work. M.E.S. provided technical guidance and feedback that led to the leakage post-selection analysis. K.M.R., G.O.S., and M.E.S. acquired funding and provided overall project supervision. K.M.R. and R.B.-K. jointly supervised P.C.W. All authors reviewed and approved the final manuscript. A LLM was used to assist in generating plotting code and verify the error decomposition expressions in the appendix.}

\section*{Acknowledgments}
The authors would like to thank Jordan Hines, Juan Gonzalez De Mendoza, and Ashe Miller for helpful discussions; Stefan Seritan for \texttt{pyGSTi} assistance; Robert Rood for contributions to measurement software; Michael Gingras, Kevin Grossklaus, Hannah Stickler, and the entire fabrication team at MIT Lincoln Laboratory for contributions to device fabrication and process development; and Steve Weber for program assistance.   

The research is based upon work supported by the Office of the Director of National Intelligence (ODNI), Intelligence Advanced Research Projects Activity (IARPA), specifically the ELQ program;  U.S. Department of Energy, Office of Science, National Quantum Information Science Research Centers, Quantum Systems Accelerator; and the Undersecretary of War for Research and Engineering under Air Force Contract No. FA8702-15-D-0001 or FA8702-25-D-B002. This work is supported by a collaboration between the U.S. DOE and other Agencies. Sandia National Laboratories is a multimission laboratory managed and operated by National Technology \& Engineering Solutions of Sandia, LLC, a wholly owned subsidiary of Honeywell International Inc., for the U.S. Department of Energy’s National Nuclear Security Administration under contract DE-NA0003525. This paper describes objective technical results and analysis. The views and conclusions contained herein are those of the authors and should not be interpreted as necessarily representing the official policies or endorsements, either expressed or implied, of the ODNI, IARPA, the Department of Energy, the Under Secretary of Defense for Research and Engineering, or the U.S. Government. The U.S. Government is authorized to reproduce and distribute reprints for Governmental purposes notwithstanding any copyright annotation thereon.

\bibliographystyle{unsrturl}
\bibliography{QIerrgen}

\onecolumn\newpage
\appendix
\section{Constructing and interpreting error generator equivalence classes}\label{appendix:equivalence_classes}
Here, we identify the error generator equivalence classes discussed in Section~\ref{sec:MCM_error_gens} for a single-qubit MCM. As we stated in the main text, we assume a linearized, small error regime where $e^L \approx I + L.$ In this first-order approximation, the total deviation $\Delta \mathcal{Q}$ can be expressed as a linear combination of 240 deviations $\Lambda^{(L_i)}$, each generated by an EEG $L_i$ [see Eq.~\eqref{eq:sum_of_EEGs}], with the EEG strengths $\epsilon_i$ serving as the coefficients:
\begin{equation}
\begin{split}
   \Delta \mathcal{Q}  &\equiv  \mathcal{Q} -\bar{\mathcal{Q}}\\ & \approx \mathcal{I}(I+L)-\mathcal{I}(I)= \mathcal{I}(L) \\
    &= \mathcal{I}\left( \sum_i \epsilon_iL_i\right)=  \sum_i\epsilon_i\mathcal{I}\left( L_i\right)\\
    &=\sum_{i=1}^{240} \epsilon_i \Lambda^{(L_i)}
\label{eq:sum_of_EEGs}
\end{split}
\end{equation}
$\mathcal{I}(\cdot)$ is the error-process-to-instrument map, introduced in Eq.~\eqref{eq:QIfromError}, which is defined for each MCM outcome $c$ as:  

\begin{equation}
\mathcal{I}_c(\cdot) \equiv (I \otimes  \langle\!\langle c| )\cdot \text{CNOT}(I\otimes |0\rangle\!\rangle) = \mathcal{Q}_c.
\end{equation}
By identifying error generator equivalence classes, we can express $\Delta \mathcal{Q}$ as a linear combination of 28 \textit{elementary deviations} $\Lambda^{(j)}$, which are linearly independent in the QI parameter space [Eq.~\eqref{eq:FOMGI_and_elementary_deviation}]. 

\begin{equation}
   \Delta \mathcal{Q}  =\sum_{i=1}^{28} \lambda_j \Lambda^{(j)}.
\label{eq:FOMGI_and_elementary_deviation}
\end{equation}
Each error coefficient $\lambda_i$ corresponds to the strength of a single effective MCM error mechanism (e.g., weakness or pure readout error), although it is composed of many two-qubit error strengths $\epsilon_i$ that are interchangeable under the MCM gauge.\footnote{For a single-qubit MCM, we can largely ignore that multiple two-qubit processes can generate the same single-qubit dynamics, because the two-qubit error process (orange splat, Fig.~\ref{fig:auxiliary_pic}) is a convenient construction with limited physical meaning. More generally, however, the observable error mechanisms in the QI can realistically be driven by multiple underlying error processes. Examples of such cases include: parity checks treated as MCMs, and single-qubit measurements implemented using terminating measurements in systems without native MCM capabilities.} \added{In the linearized regime,} \deleted{we refer to} these linear combinations $\lambda_j=\sum _i\alpha_i\epsilon_i$ \added{are} \deleted{as} \textit{first-order MCM-gauge-invariant (FOMGI)} quantities. \added{This error decomposition is possible for all quantum instruments describing a single-qubit MCM (and although out of the scope of this work, should be possible more generally). However, outside the small-error regime second-order effects may become significant, leading to the association of $\lambda_j$ with FOMGI quantities breaking down.}

\subsection{Derivation of elementary deviations}\label{appendix:derivation}
We construct the elementary deviations by studying the actions of EEGs in the single-qubit picture, and identifying the elementary deviations as their unique actions on the QI. Each EEG is a combination of (1) a symmetric term with action $P\rho Q + Q \rho P$  \textit{or} an anti-symmetric term with action $i(P\rho Q - Q \rho P)$\footnote{These terms are often referred to in the literature as \textit{(anti-)symmetrized Choi units} \cite{blume-kohout2022a}.}, and (2) a symmetric correction term of the form  $I\rho R + R \rho I$ that ensures the error generator is trace-preserving \cite{blume-kohout2022a}.

This fact suggests that we should consider $P\rho'Q+Q\rho'P$ and $i(P\rho'Q-Q\rho'P)$ in turn. Here, $P=P^{(D)} \otimes P^{(V)}$ and $Q=Q^{(D)} \otimes Q^{(V)}$ are tensor products of single-qubit Pauli operators acting on the physical ($D$) and virtual ($V$) qubits. For simplicity, we only include the qubit labels where the meaning might not otherwise be clear. Assuming the initial state of the physical qubit is $\rho$, the joint physical+virtual state $\rho'$ after the CNOT (blue dashed line, Fig.~\ref{fig:auxiliary_pic}(b)) is:

\begin{equation}
\begin{split}
    \rho' &= \text{CNOT}_{DV}(\rho \otimes |0\rangle\!\langle 0|)  \text{CNOT}_{DV} \\ 
    &= \sum_{a,b} \langle a| \rho|b\rangle  |a\rangle\!\langle b| \otimes |a\rangle\!\langle b|.
\end{split}
\label{jointstate}
\end{equation}
To recover the action of these terms solely on the physical qubit when the MCM outcome is $c$, we project the virtual qubit onto $|c\rangle \!\langle c|$. This allows us to define the deviation $\Lambda^{P,Q, \mathrm{sym}}=\{\Lambda^{P,Q, \mathrm{sym}}_0, \Lambda^{P,Q, \mathrm{sym}}_1\}$ derived from the symmetric term $P \rho Q + Q \rho P$. Specifically, we have:

\begin{equation}
\begin{split}
    \Lambda^{P,Q,\mathrm{sym}}_c[\rho] & = \text{Tr}_V\left[(I \otimes |c\rangle\!\langle c|)(P\rho'Q + Q\rho' P)(I \otimes |c\rangle\!\langle c|)\right] \\
    &= \sum_{a,b}\bigg( \langle a| \rho|b\rangle \cdot \langle c| P^{(V)}|a \rangle \cdot \langle b | Q^{(V)}|c\rangle \cdot P^{(D)}|a\rangle\!\langle b| Q^{(D)} \\
    &\phantom{\sum_{a,b}\bigg( } +  \langle a| \rho|b\rangle \cdot \langle c| Q^{(V)}|a \rangle \cdot \langle b | P^{(V)}|c\rangle \cdot  Q^{(D)}|a\rangle\!\langle b| P^{(D)}\bigg)
\label{eq:eeg_on_state_sym}
\end{split}
\end{equation}
Similarly, we can express the deviation $\Lambda^{P,Q, \mathrm{ anti}}=\{\Lambda^{P,Q, \mathrm{anti}}_0, \Lambda^{P,Q, \mathrm{anti}}_1\}$ derived from the antisymmetric term $i(P \rho Q - Q \rho P)$ as follows:

\begin{equation}
\begin{split}
    \Lambda^{P,Q,\mathrm{anti}}_c[\rho]  &= \text{Tr}_V \left[  (I \otimes |c\rangle\!\langle c|)i(P\rho'Q - Q\rho' P)(I \otimes |c\rangle\!\langle c|)\right] \\
    &= 
    i\sum_{a,b}\bigg( \langle a| \rho|b\rangle \cdot \langle c| P^{(V)}|a \rangle \cdot \langle b | Q^{(V)}|c\rangle \cdot P^{(D)}|a\rangle\!\langle b| Q^{(D)} \\
    &\phantom{i\sum_{a,b}\bigg(} -  \langle a| \rho|b\rangle \cdot \langle c| Q^{(V)}|a \rangle \cdot \langle b | P^{(V)}|c\rangle \cdot  Q^{(D)}|a\rangle\!\langle b| P^{(D)}\bigg)
\label{eq:eeg_on_state_antisym}
\end{split}
\end{equation}

By studying these equations, we can identify $32$ unique deviations. These elementary deviations correspond to the $32$ parameters required to parameterize the QI with the TP constraint relaxed. First, we observe that the total action can initially be either symmetric [Eq.~\eqref{eq:eeg_on_state_sym}] or anti-symmetric [Eq.~\eqref{eq:eeg_on_state_antisym}]. Next, we study the action of the Pauli operators on $\rho'$. An $X$ Pauli operator flips a bit. There are four possible bits that can be uniquely flipped in Eqs.~\eqref{eq:eeg_on_state_sym} and \eqref{eq:eeg_on_state_antisym}, resulting in $8$ different bit flip patterns. The factor of $2$ is due to the symmetry of the expression in Eqs.~\eqref{eq:eeg_on_state_sym} and~\eqref{eq:eeg_on_state_antisym}. A $Z$ operator introduces a relative phase between the error processes corresponding to the $|0\rangle$ and $|1\rangle$ measurement outcomes. Consequently, the output is either maximally correlated or completely uncorrelated with the readout bit. A $Y$ Pauli operator flips a bit, introduces the phase flip, \textit{and} flips the initial symmetry of the expression. Ultimately, this yields a total of $2 \times 8 \times 2 = 32$ elementary deviations.  

We enforce trace preservation by adding a correction term to all deviations that are not intrinsically TP. This correction term takes the same form as the correction term present in the EEGs: $-\frac{1}{2}\{ \{P,Q\}, \rho' \}$ for all symmetric terms and $\frac{1}{2}\{ [P,Q], \rho' \}$ for all anti-symmetric terms. This step also eliminates four deviations which are non-TP. The coefficients of the $28$ remaining deviations, the FOMGI quantities, parameterize the QI \textit{with} the TP constraint. 

\subsection{Construction of FOMGI quantities: examples}\label{appendix:examples}
To illustrate the construction of FOMGI quantities, we revisit the examples we used in the main text to motivate this error decomposition. We previously showed that $S_{IX}$ and $S_{IY}$ produce indistinguishable error processes in the QI. As such, the error rates $s_{ix}$ and $s_{iy}$ associated with these EEGs must contribute to the same FOMGI quantity, which we denote $s^{\mathrm{prep}}$ (see Appendix~\ref{appendix:classification} for the motivation behind this nomenclature). We can also explicitly see this by applying our rules for the action of various Pauli operators in Eqs.~\eqref{eq:eeg_on_state_sym} and \eqref{eq:eeg_on_state_antisym}. The base term of $S_{IY}$ has two $Y$ operators, whose effect on the symmetry and relative phase cancel out, producing an identical action on the state $\rho'.$ Using the same approach, we can show that the coefficients of $S_{ZX}$, $S_{ZY}$, $A_{IXZY}$, and $A_{IYZX}$ all contribute to $s^{\mathrm{prep}}$: 
\begin{equation}
    s^{\mathrm{prep}}=s_{ix}+s_{iy}+s_{zx}+s_{zy}+a_{ixzy}-a_{iyzx}
    \label{eq:FOMGI1}
\end{equation}
The elementary deviation corresponding to this FOMGI quantity is shown on the right side of Fig.~\ref{fig:errorgen_to_QI}(a). 

Next, we construct the FOMGI quantity that includes $H_{ZY}$, which has an action shown in Fig.~\ref{fig:errorgen_to_QI}(c). Using Eqs.~\eqref{eq:eeg_on_state_sym} and~\eqref{eq:eeg_on_state_antisym}, we can determine that correlation and active EEGs can produce the same error process. Ultimately, the EEG terms that contribute to this FOMGI quantity are as follows: 

\begin{itemize}
    \item A single Hamiltonian error generator $H_{ZY}$.
    \item The base terms of $C_{IXZZ}$ and $C_{IZZX}$.
    \item The correction terms of $C_{XIXX}$, $C_{YIYX}$, $A_{XYXZ}$, and $A_{YYYZ}$.
    \item The base \textit{and} correction terms of $C_{ZIZX}$ and $A_{ZYZZ}$ resulting in these EEGs contributing twice as much to the FOMGI quantity.  
\end{itemize}
Ultimately, the full FOMGI quantity, which we denote $w_0$, is 
\begin{equation}
\begin{split}
    w_0  &= h_{zy} + a_{iyzi} - a_{xyxz} - a_{yyyz} - 2a_{zyzz} \\ &+ c_{ixzz} - c_{izzx} - c_{xixx} - c_{yiyx} - 2c_{zizx},
    \label{eq:FOMGI2}
\end{split}
\end{equation}
and the elementary deviation corresponds to the first-order effects in the deviation depicted on the right of Fig.~\ref{fig:errorgen_to_QI}(c). 

\section{A taxonomy of MCM errors} \label{appendix:classification}

\begin{figure}
    \centering
    \includegraphics[width=0.75\linewidth]{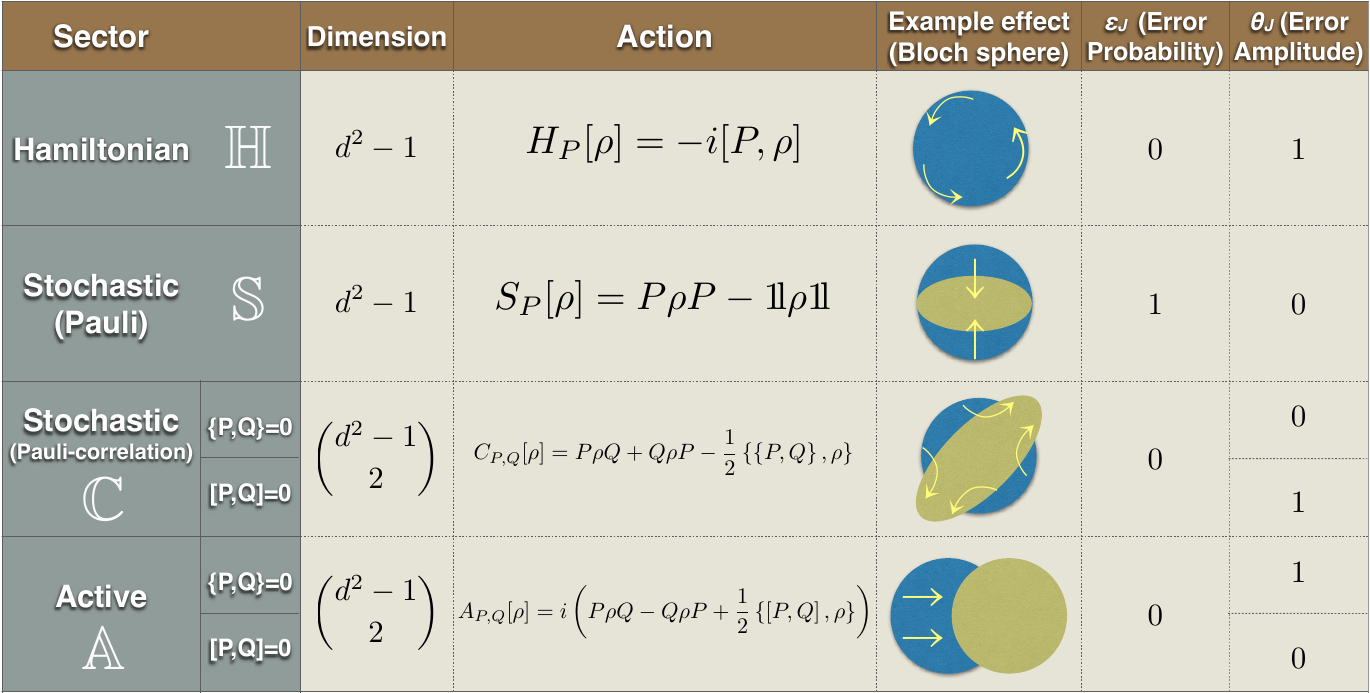}
    \caption{\added{Elementary error generators (EEGs) are divided into four sectors: Hamiltonian (H), Pauli-stochastic (S), Pauli-Correlation (C), and active (A). The table shows the mathematically representation of the action of each sector and illustrates how a representative single-qubit EEG from each sector transforms the Bloch sphere. Figure reproduced with permission from Ref.~\cite{blume-kohout2022a}.}}
    \label{fig:taxonomy_original}
\end{figure}

The original error generator framework is useful because it provides the ability to classify EEGs according to physically meaningful sectors and weights \added{(see Fig`\ref{fig:taxonomy_original})}. We suggest a similar classification below. First we observe that FOMGI quantities can be constructed so that no quantity contains both Pauli-stochastic and Hamiltonian error rates. This suggests that a sector-based classification mirroring the standard error generator sectors may be useful. While it is worth noting that all FOMGI quantities containing Pauli-stochastic and Hamiltonian terms also include active or correlation terms, we find that their action is most easily interpreted using the Hamiltonian or Pauli-stochastic term. Additionally,  we can consider the qubit(s) on which the EEGs act. Certain FOMGI quantities contain terms that effectively act on only the data qubit. These two observations allow us to tease out classes of MCM errors, each denoted by a capital letter in the same fashion as the error generator framework for gates.

\begin{itemize}
    \item \textbf{Pauli-stochastic ($S$)} FOMGI quantities [e.g., Eq.~\eqref{eq:FOMGI1}] contain at least one Pauli-stochastic error rate. These FOMGI quantities quantify the strength of probabilistic Pauli errors,  which are the only error rates present in uniform stochastic instruments. These errors capture effective depolarization and pure readout error. There are three $S$ errors, which are reported in Table~\ref{tab:stochastic}.
    \item \textbf{Hamiltonian ($H$)} FOMGI quantities [e.g., Eq.~\eqref{eq:FOMGI2}] contain at least one Hamiltonian error rate. As such, these errors can be treated as being generated by unitary dynamics on the data$+$virtual qubit system. 
    \begin{itemize}
        \item \textbf{Axis rotations ($R$)}: the quantum state is rotated around the measurement axis or state (re)preparation axis. These errors can be generated by a weight-one Hamiltonian error acting on the data qubit before or after the MCM. The combination of measurement-axis-only and state-preparation-axis-only rotations produce measurement-axis misalignment, where the MCM projects onto a slightly rotated eigenstate. In total, there are six such errors, with two acting solely on the measurement axis ($R^{\mathrm{meas}}$) and four on the state preparation axis ($R^{\mathrm{prep}}$). We detail the full list of $R$ errors in Table~\ref{tab:rotation}.
        \item \textbf{Unitary weakness ($W$)}: the quantum state of the register is not completely collapsed into an eigenstate of the measured observable, leaving anticommuting observables not completely annihilated. In the gadget picture (Fig.~\ref{fig:auxiliary_pic}), incomplete collapse occurs if the CNOT gate does not properly rotate the virtual qubit conditional on the state of the physical qubit. Here, the CNOT error process can be treated as a unitary error process. There are four $W$ errors, which we detail in Table~\ref{tab:coherence}. 
    \end{itemize}
    \item \textbf{Pauli-correlation/Active ($C$/$A$)} FOMGI quantities that contain only Pauli-correlation and active error rates. Just as correlation and active EEGs rarely appear in theoretical models for single-qubit gate errors \cite{blume-kohout2022a}, C/A FOMGI quantities are generally absent from physical models of single-qubit MCM error.  $T_1$ decay processes, i.e., amplitude damping from $|0\rangle$ to $|1\rangle$, remain a key exception. 
    \begin{itemize}
        \item \textbf{Active ($A$) errors}: these errors combine with the Pauli-stochastic errors above to drive amplitude damping and measurement bias processes. We refer to these errors as  ``active'' for two reasons: (1) they can be produced by a weight-one active error generator acting either before or after the measurement and (2) they are quantitatively similar to the active errors in gates. There are three $A$ errors, which we detail in Table~\ref{tab:stochastic_counterparts}.
        \item \textbf{$\rho$-dependent axis rotations ($\tilde{R}$)}: the quantum state is rotated around the measurement axis or state (re)preparation axis. Here, the rotation depends on the input state $\rho$. There are eight of these errors, four of which act on the measurement axis and four on the state preparation axis. These errors ``modify'' $R$ errors to produce more exotic effects. We detail the full list of $\tilde{R}$ errors in Table~\ref{tab:rotation_counterparts}.
        \item \textbf{Non-unitary weakness ($\tilde{W}$)}: the quantum state of the register is not completely collapsed into an eigenstate of the measured observable, leaving anticommuting observables not completely annihilated. In the gadget picture (Fig.~\ref{fig:auxiliary_pic}), incomplete collapse occurs if the CNOT gate does not properly rotate the virtual qubit conditional on the state of the physical qubit. Unlike $W$ errors, the CNOT error process \textit{cannot} be treated as a unitary error process. There are four $\tilde{W}$ errors,  which we detail in Table~\ref{tab:weak_counterparts}. 
    \end{itemize}
\end{itemize}
Like the original error generator framework for gates, we often find that physical error processes are not driven by a single elementary deviation. We find that it is sometimes more informative to report the error strengths associated with the sums and differences of elementary deviations (e.g, Figs.~\ref{fig:error_mechanisms} and \ref{fig:error_decomposition}) rather than single elementary deviation. For example, $T_1$ decay (amplitude damping from $|1\rangle$ to $|0\rangle$) during readout is a linear combination of a Pauli-stochastic elementary deviation $S^{\mathrm{meas}}$ and an active elementary deviation $A^{\mathrm{meas}}$. The deviation is 

\begin{equation}
\begin{split}
          \Gamma_{1 \rightarrow 0}^{\mathrm{meas}}[\rho] &= 2S^{\mathrm{meas}}[\rho]-A^{\mathrm{meas}}[\rho] \\ &= \{4\rho_{11}|0\rangle\!\langle 0|, -4\rho_{11}|1\rangle\!\langle 1|\}.
\end{split}
\end{equation}

We also choose to report axis rotations and phase error on residual coherence such that the error strengths capture the error on a measurement outcome $c$. For example, the deviation for an axis rotation around $x$-axis on output state for a measurement outcome $c$ is 

\begin{equation}
R_{x,c=0}^{\mathrm{prep}}[\rho] = \frac{1}{2}(R_{x,\mathrm{ind}}^{\mathrm{prep}}[\rho]-R_{x,\mathrm{dep}}^{\mathrm{prep}}[\rho]) = \{\rho_{00}Y, \mathbf{0}\}
\end{equation}
Due to how we construct these deviations, they remain linearly independent in the QI parameter space. 

\begin{table}[H]
    \centering
    \begin{tabular}{@{} p{0.8cm} p{0.8cm}p{3.8cm}p{5.6cm}p{4.5cm}@{}l}
    \toprule
    Label ($\lambda$) & Rep. term& FOMGI quantity &Unit action \newline ($\Lambda[\rho] = \{\Lambda_0[\rho], \Lambda_1[\rho] \}$)  & Unit change in probs. ($\Delta p_c = \text{Tr}(\Lambda_c[\rho])$) \\
    \midrule
        $s^{\mathrm{meas}}$ & $s_{xx}$ & \parbox[t!]{3.8cm}{$s_{xx} + s_{yx} + s_{xy} + s_{yy}-2c_{xxyy} + 2c_{xyyx}$} &  \parbox[t!]{5.6cm}{\begin{align*}
          S^{\mathrm{prep}}[\rho] = & \{(\rho_{00}-\rho_{11})|0\rangle\!\langle 0|,\\ &\{(\rho_{00}-\rho_{11})|1\rangle\!\langle 1|\}
          \end{align*}}&\parbox[t!]{4.5cm}{\begin{align*}
          \Delta p_0 &= (\rho_{00}-\rho_{11})\\ 
          \Delta p_1 &= (\rho_{00}-\rho_{11})
          \end{align*}}\\ 
    \midrule
         $s^{\mathrm{prep}}$ & $s_{xi}$ & \parbox[t!]{3.8cm}{$s_{xi}+s_{yi}+s_{xz}+ s_{yz} +2 a_{xiyz} + 2a_{xzyi}$} & \parbox[t!]{5cm}{\begin{align*}
         S^{\mathrm{prep}}[\rho] =  \{-\rho_{00} Z, \rho_{11}Z\} 
          \end{align*}}&
           $\Delta p_0 = \Delta p_1 = 0$ \\
    \midrule
         $s^{\mathrm{read}}$ & $s_{ix}$ & \parbox[t!]{3.8cm}{$s_{ix} + s_{iy} + s_{zx} + s_{zy}+2a_{ixzy} - 2a_{iyzx}$} & \parbox[t!]{5.6cm}{\begin{align*}
          S^{\mathrm{read}}[\rho] = & \{-\rho_{00} |0\rangle\!\langle 0|+\rho_{11} |1\rangle\!\langle 1|,\\ &\{\rho_{00} |0\rangle\!\langle 0|-\rho_{11} |1\rangle\!\langle 1|\}
          \end{align*}} &\parbox[t!]{4.5cm}{\begin{align*}
          \Delta p_0 &= -\rho_{00}+\rho_{11}\\ 
          \Delta p_1 &= \rho_{00}-\rho_{11}
          \end{align*}}\\ 
    \bottomrule
    \end{tabular}
    \caption{\textbf{Pauli-stochastic ($S$) errors}: these errors capture effective depolarization during measurement ($s^{\mathrm{meas}}$), depolarization during re-preparation ($s^{\mathrm{prep}}$), and pure readout error ($s^{\mathrm{read}}$). In each case, we report a representative term in the FOMGI quantity in addition to the label, which highlights how we determine the action of a given elementary deviation. To illuminate the structure of this decomposition, we show the unit action ($\Lambda[\rho] =\{\Lambda_0[\rho], \Lambda_1[\rho] \}$) and unit change in probabilities ($\Delta p_c = \text{Tr}(\Lambda_c[\rho])$). For a elementary deviation $\Lambda$ acting with strength $\lambda$, the output state is $\lambda\Lambda_c[\rho]$ and the change in measurement probabilities is $\text{Tr}(\lambda\Lambda_c[\rho])$.}
    \label{tab:stochastic}
\end{table}

\begin{table}[H]
    \centering
    \begin{tabular}{@{} p{0.8cm} p{0.8cm}p{4cm} p{5.4cm}p{4.5cm}@{}l}
    \toprule
    Label ($\lambda$) & Rep. term& FOMGI quantity &Unit action \newline ($\Lambda[\rho] = \{\Lambda_0[\rho], \Lambda_1[\rho] \}$)  & Unit change in probs. ($\Delta p_c = \text{Tr}(\Lambda_c[\rho])$) \\
    \midrule
        $a^{\mathrm{meas}}$ & $a_{xxyx}$ & \parbox[t!]{4cm}{ $a_{xxyx}+ a_{xxxy} + a_{xyyy} + a_{yxyy}$} & \parbox[t!]{5.4cm}{\begin{align*}
          A^{\mathrm{meas}}[\rho] =&  \{-2(\rho_{00}+\rho_{11})|0\rangle\!\langle 0|,\\ & 2(\rho_{00}+\rho_{11})|1 \rangle\!\langle 1|\} 
          \end{align*}} & \parbox[t!]{4.5cm}{\begin{align*}
          \Delta p_0 &= -2(\rho_{00}+\rho_{11})\\ 
          \Delta p_1 &= 2(\rho_{00}+\rho_{11})
          \end{align*}}\\ 
    \midrule
         $a^{\mathrm{prep}}$ & $a_{xiyi}$ & \parbox[t!]{4cm}{$a_{xiyi} + a_{xzyz} + c_{xixz} + c_{yiyz}$} & \parbox[t!]{5cm}{\begin{align*}
         A^{\mathrm{prep}}[\rho] =  \{-2\rho_{00}Z, -2\rho_{11}Z\} 
          \end{align*}} &
          $\Delta p_0 = \Delta p_1 = 0$\\  
    \midrule
         $a^{\mathrm{read}}$ & $a_{ixiy}$ & \parbox[t!]{4cm}{ $a_{ixiy} + a_{zxzy} + c_{ixzx} + c_{iyzy}$} & \parbox[t!]{5.4cm}{\begin{align*}
          A^{\mathrm{read}}[\rho] =&  \{-2\rho_{00}|0\rangle\!\langle 0|-2\rho_{11}|1\rangle\!\langle 1|,\\
           &2\rho_{00}|0\rangle\!\langle 0|+ 2\rho_{11}|1 \rangle\!\langle 1|\} 
          \end{align*}} & \parbox[t!]{4.5cm}{\begin{align*}
          \Delta p_0 &= -2(\rho_{00}+\rho_{11})\\ 
          \Delta p_1 &= 2(\rho_{00}+\rho_{11})
          \end{align*}}\\ 
    \bottomrule
    \end{tabular}
    \caption{\textbf{Active ($A$) errors}: in conjunction with Pauli-stochastic errors, these generate outcome-dependent bias  (e.g., $2S^{\mathrm{prep}}-A^{\mathrm{prep}}$ generates amplitude damping on the re-prepared state of the mid-circuit measurement). A positive error strength indicates a bias towards $|1\rangle$; a negative error strength indicates a bias toward $|0\rangle.$ In each case, we report a representative term in the FOMGI quantity in addition to the label, which highlights how we determine the action of a given elementary deviation. To illuminate the structure of this decomposition, we show the unit action ($\Lambda[\rho] =\{\Lambda_0[\rho], \Lambda_1[\rho] \}$) and unit change in probabilities ($\Delta p_c = \text{Tr}(\Lambda_c[\rho])$). For a elementary deviation $\Lambda$ acting with strength $\lambda$, the output state is $\lambda\Lambda_c[\rho]$ and the change in measurement probabilities is $\text{Tr}(\lambda\Lambda_c[\rho])$.}
    \label{tab:stochastic_counterparts}
\end{table}

\begin{table}[H]
    \centering
    \begin{tabular}{@{} p{1cm} p{1cm}p{5cm} p{5cm}p{3.5cm}@{}l}
    \toprule
     Label ($\lambda$) & Rep. term& FOMGI quantity & Unit action ($\Lambda[\rho] = \{\Lambda_0[\rho], \Lambda_1[\rho] \}$)  & Unit change in probs. ($\Delta p_c = \text{Tr}(\Lambda_c[\rho])$) \\
    \midrule
          $r_{x}^{\mathrm{meas}}$ & $h_{xx}$ & \parbox[t!]{5cm}{$h_{xx}-h_{yy}- a_{xxzz} + a_{yyzz} + c_{izxy} + c_{izyx} + c_{xyzi} + c_{yxzi}$} & 
          \parbox[t!]{5cm}{\begin{align*}
          R_x^{\mathrm{meas}}[\rho] =& \{-i\left(\rho_{01}-\rho_{10}\right)  |0 \rangle\!\langle 0|,\\
           & i\left(\rho_{01} -\rho_{10}\right)|1 \rangle\!\langle 1|\} 
          \end{align*}} & \parbox[t!]{3.5cm}{\begin{align*}
          \Delta p_0 &= -i\left(\rho_{01}-\rho_{10}\right)\\ 
          \Delta p_1 &= i\left(\rho_{01} -\rho_{10}\right) 
          \end{align*}}\\ 
    \midrule
         $r_{y}^{\mathrm{meas}}$ & $h_{yx}$ & \parbox[t!]{5cm}{$h_{yx}+h_{xy}-a_{xyzz} - a_{yxzz} - c_{izxx} + c_{izyy} - c_{xxzi} + c_{yyzi}$} & \parbox[t!]{5cm}{\begin{align*}
          R_y^{\mathrm{meas}}[\rho] =& \{(\rho_{01}+\rho_{10}) |0 \rangle\!\langle 0|,\\
          & -(\rho_{01} + \rho_{10}) |1 \rangle\!\langle 1| \}
          \end{align*}} & \parbox[t!]{3.5cm}{\begin{align*}
          \Delta p_0 &= \left(\rho_{01}+\rho_{10}\right)\\ 
          \Delta p_1 &= -\left(\rho_{01} +\rho_{10}\right) 
          \end{align*}}\\ 
    \midrule
        $r_{x,\mathrm{ind}}^{\mathrm{prep}}$ & $h_{xi}$  & \parbox[t!]{5cm}{$h_{xi}+a_{ixyy} - a_{iyyx} + a_{izxz} - 2 a_{xizz} - 2a_{xzzi} + c_{xxzy} - c_{xyzx} - c_{yizi} - c_{yzzz}$} & \parbox[t!]{5cm}{\begin{align*}
         R_{x,\mathrm{ind}}^{\mathrm{prep}}[\rho] = \{-\rho_{00}Y, 
           \rho_{11}Y\} 
          \end{align*}} & 
          $\Delta p_0 = \Delta p_1 =0$\\ 
    \midrule
         $r_{x,\mathrm{dep}}^{\mathrm{prep}}$ & $h_{xz}$ & \parbox[t!]{5cm}{$h_{xz}+a_{izxi} - 2a_{xizi} - a_{xxzx} - a_{xyzy} - 2a_{xzzz} + c_{ixyx} + c_{iyyy} - c_{yizz} - c_{yzzi}$} & \parbox[t!]{5cm}{\begin{align*}
          R^{x,\mathrm{dep}}_{\mathrm{prep}}[\rho] = \{-\rho_{00} Y, -\rho_{11} Y\}
          \end{align*}}  & $\Delta p_0 = \Delta p_1 =0$\\ 
    \midrule
         $r_{y,\mathrm{ind}}^{\mathrm{prep}}$ & $h_{yi}$ & \parbox[t!]{5cm}{$h_{yi}-a_{ixxy} + a_{iyxx} + a_{izyz} - 2a_{yizz} - 2 a_{yzzi} + c_{xizi} + c_{xzzz} + c_{yxzy} - c_{yyzx}$} & \parbox[t!]{5cm}{\begin{align*}
          R_{y,\mathrm{ind}}^{\mathrm{prep}}[\rho] = \{-\rho_{00}X, \rho_{11}X\}
          \end{align*}} &  $\Delta p_0 = \Delta p_1 =0$\\ 
    \midrule
         $r_{y,\mathrm{dep}}^{\mathrm{prep}}$ &$h_{yz}$ & \parbox[t!]{5cm}{$h_{yz}+ a_{izyi} - 2a_{yizi} - a_{yxzx} - a_{yyzy} -2 a_{yzzz} - c_{ixxx} - c_{iyxy} + c_{xizz} + c_{xzzi}$} & \parbox[t!]{5cm}{\begin{align*}
          R_{y,\mathrm{dep}}^{\mathrm{prep}}[\rho] &= \{-\rho_{00}X, -\rho_{11}X\}
          \end{align*}} & $\Delta p_0 = \Delta p_1 =0$\\ 
    \bottomrule
    \end{tabular}
    \caption{\textbf{Axis rotations}: the quantum state rotates around the measurement axis or state preparation axis. $R_{x}^{\mathrm{meas}}$ and $R_{y}^{\mathrm{meas}}$ produce measurement-axis-only rotations about the $x$- and $y$-axes, respectively. The remaining four quantities, $R_{x,\mathrm{ind}}^{\mathrm{prep}}$, $R_{x,\mathrm{dep}}^{\mathrm{prep}}$, $R_{y,\mathrm{ind}}^{\mathrm{prep}}$, and $R_{y,\mathrm{dep}}^{\mathrm{prep}}$, produce state-preparation-only-axis rotations. Among the post-MCM rotations, $R_{x,\mathrm{ind}}^{\mathrm{prep}}$ and $R_{y,\mathrm{ind}}^{\mathrm{prep}}$ are outcome-independent, meaning the direction of rotation does not depend on the outcome, while $R_{x,\mathrm{dep}}^{\mathrm{prep}}$ and $R_{y,\mathrm{dep}}^{\mathrm{prep}}$ are outcome-dependent. In each case, we report a representative term in the FOMGI quantity in addition to the label, which highlights how we determine the action of a given elementary deviation. To illuminate the structure of this decomposition, we show the unit action ($\Lambda[\rho] =\{\Lambda_0[\rho], \Lambda_1[\rho] \}$) and unit change in probabilities ($\Delta p_c = \text{Tr}(\Lambda_c[\rho])$). For a elementary deviation $\Lambda$ acting with strength $\lambda$, the output state is $\lambda\Lambda_c[\rho]$ and the change in measurement probabilities is $\text{Tr}(\lambda\Lambda_c[\rho])$.}
    \label{tab:rotation}
\end{table}

\begin{table}[H]
    \centering
    \begin{tabular}{@{} p{0.8cm} p{0.8cm}p{5cm} p{5cm}p{3.5cm}@{}l}
    \toprule
     Label ($\lambda$) & Rep. term& FOMGI quantity & Unit action ($\Lambda = \{\Lambda_0, \Lambda_1 \}$)  & Unit change in probs. ($\Delta p_c = \text{Tr}(\Lambda_c[\rho])$) \\
        \midrule
         $\tilde{r}_{x}^{\mathrm{meas}}$&$c_{xizy}$ & \parbox[t!]{5cm}{$a_{ixxi}-a_{iyyi}+a_{xzzx}-a_{yzzy}-c_{ixyz}-c_{iyxz}+c_{xizy}+c_{yizx}$} & \parbox[t!]{5cm}{\begin{align*}
          \tilde{R}_x^{\mathrm{meas}}[\rho] =& \{-i(\rho_{01}-\rho_{10}) |1\rangle\!\langle 1|,\\ 
          & i(\rho_{01}-\rho_{10}\rangle) |0\rangle\!\langle 0| \}
          \end{align*}} & \parbox[t!]{3.5cm}{\begin{align*}
          \Delta p_0 &= -i\left(\rho_{01}-\rho_{10}\right)\\ 
          \Delta p_1 &= i\left(\rho_{01} -\rho_{01}\right) 
          \end{align*}} \\ \midrule
          $\tilde{r}_{y}^{\mathrm{meas}}$&$c_{ixxz}$ &\parbox[t!]{5cm}{$a_{ixyi}+a_{iyxi}+a_{xzzy}+a_{yzzx}+c_{ixxz}-c_{iyyz}-c_{xizx}+c_{yizy}$} &  \parbox[t!]{5cm}{\begin{align*}
          \tilde{R}_y^{\mathrm{meas}}[\rho] =&\{(\rho_{01}+ \rho_{10})|1 \rangle\!\langle 1|,  \\
          & -(\rho_{01}+ \rho_{10})|0 \rangle\!\langle 0|\}
          \end{align*}}
          & \parbox[t!]{3.5cm}{\begin{align*}
          \Delta p_0 &= \left(\rho_{01}+\rho_{10}\right)\\ 
          \Delta p_1 &= -\left(\rho_{01} +\rho_{10}\right) 
          \end{align*}}\\  \midrule
         $\tilde{r}_{x,z}^{\mathrm{meas}}$& $c_{xzzy}$ &\parbox[t!]{5cm}{$a_{ixxz} - a_{iyyz} + a_{xizx} - a_{yizy} - c_{ixyi} - c_{iyxi} + c_{xzzy} + c_{yzzx}$} & \parbox[t!]{5cm}{\begin{align*}
          \tilde{R}_{x,z}^{\mathrm{meas}}[\rho] =& \{-i(\rho_{01}-\rho_{10})|1\rangle\langle1|, \\
          & -i(\rho_{01}-\rho_{10})|0\rangle\langle0|\}
          \end{align*}}
          & \parbox[t!]{3.5cm}{\begin{align*}
          \Delta p_0 &= -i\left(\rho_{01}-\rho_{10}\right)\\ 
          \Delta p_1 &= -i\left(\rho_{01} -\rho_{01}\right) 
          \end{align*}} \\ \midrule
         $\tilde{r}_{y,z}^{\mathrm{meas}}$& $c_{ixxi}$ &\parbox[t!]{5cm}{$a_{ixyz} + a_{iyxz} + a_{xizy} + a_{yizx} + c_{ixxi} - c_{iyyi} - c_{xzzx} + c_{yzzy}$} &  \parbox[t!]{5cm}{\begin{align*}
          \bar{R}_{y,z}^{\mathrm{meas}}[\rho] =&\{(\rho_{01}+\rho_{10})|1\rangle\langle1|, \\
          & (\rho_{01}+\rho_{10})|0\rangle\langle0|\}
          \end{align*}}
          & \parbox[t!]{3.5cm}{\begin{align*}
          \Delta p_0 &= \left(\rho_{01}+\rho_{10}\right)\\ 
          \Delta p_1 &= \left(\rho_{01} +\rho_{10}\right) 
          \end{align*}}\\ 
     \midrule
         $\tilde{r}_{x,\mathrm{ind}}^{\mathrm{prep}}$& $c_{xxzy}$ &\parbox[t!]{5cm}{$a_{ixxx} - a_{ixyy} + a_{iyxy} + a_{iyyx} + c_{xxzy} - c_{xyzx} - c_{yxzx} - c_{yyzy}$} & \parbox[t!]{5cm}{\begin{align*}
         \bar{R}_{x,\mathrm{ind}}^{\mathrm{prep}}[\rho] =\{\rho_{11}Y, -\rho_{00}Y\}
          \end{align*}} & $\Delta p_0 = \Delta p_1 = 0$\\ 
    \midrule
         $\tilde{r}_{x,\mathrm{dep}}^{\mathrm{prep}}$& $c_{ixxy}$ &\parbox[t!]{5cm}{$a_{xxzx} + a_{xyzy} + a_{yxzy} - a_{yyzx} + c_{ixxy} + c_{ixyx} - c_{iyxx} + c_{iyyy}$} &  \parbox[t!]{5cm}{\begin{align*}
         \tilde{R}_{x,\mathrm{dep}}^{\mathrm{prep}}[\rho] =\{-\rho_{11}Y, -\rho_{00} Y\}
          \end{align*}} & $\Delta p_0 = \Delta p_1 = 0$\\ 
    \midrule
         $\tilde{r}_{y,\mathrm{ind}}^{\mathrm{prep}}$ &$c_{xxzx}$ &\parbox[t!]{5cm}{$-a_{ixxy} - a_{ixyx} +a_{iyxx} - a_{iyyy} - c_{xxzx} - c_{xyzy} - c_{yxzy} + c_{yyzx}$} & \parbox[t!]{5cm}{\begin{align*}
          \tilde{R}_{y,\mathrm{ind}}^{\mathrm{prep}}[\rho] &= \{\rho_{11}X,-\rho_{00}X\}
          \end{align*}}& $\Delta p_0 = \Delta p_1 = 0$\\ 
    \midrule
        $\tilde{r}_{y, \mathrm{dep}}^{\mathrm{prep}}$&$c_{ixyy}$ &\parbox[t!]{5cm}{$-a_{xxzy} + a_{xyzx} + a_{yxzx} + a_{yyzy} - c_{ixxx} + c_{ixyy} - c_{iyxy} - c_{iyyx}$}& \parbox[t!]{5cm}{\begin{align*}
          \bar{R}_{y,\mathrm{dep}}^{\mathrm{prep}}[\rho] &= \{-\rho_{11}X,-\rho_{00}X\}
          \end{align*}} & $\Delta p_0 = \Delta p_1 = 0$\\  
    \bottomrule
    \end{tabular}
    \caption{\textbf{$\rho$-dependent axis rotations ($\tilde{R}$)}: these errors produce measurement-axis and state-preparation-axis rotations that are dependent on the input state. All but two $\tilde{R}$ errors (i.e., $\tilde{R}_{x,z}^{\mathrm{meas}}$ and $\tilde{R}_{y,z}^{\mathrm{meas}}$) have a counterpart $R$ error, which is $\rho$-independent. In each case, we report a representative term in the FOMGI quantity in addition to the label, which highlights how we determine the action of a given elementary deviation. In each case, we report a representative term in the FOMGI quantity in addition to the label, which highlights how we determine the action of a given elementary deviation. To illuminate the structure of this decomposition, we show the unit action ($\Lambda[\rho] =\{\Lambda_0[\rho], \Lambda_1[\rho] \}$) and unit change in probabilities ($\Delta p_c = \text{Tr}(\Lambda_c[\rho])$). For a elementary deviation $\Lambda$ acting with strength $\lambda$, the output state is $\lambda\Lambda_c[\rho]$ and the change in measurement probabilities is $\text{Tr}(\lambda\Lambda_c[\rho])$.}
    \label{tab:rotation_counterparts}
    
\end{table}

\begin{table}[H]
    \centering
    \begin{tabular}{@{} p{0.8cm} p{0.8cm}p{3.9cm} p{6.5cm}p{3.5cm}@{}l}
    \toprule
     Label ($\lambda$) & Rep. term& FOMGI quantity & Unit action ($\Lambda = \{\Lambda_0, \Lambda_1 \}$)  & Unit change in probs. ($\Delta p_c = \text{Tr}(\Lambda_c[\rho])$) \\
        \midrule
         $w_0$ & $h_{zy}$ &\parbox[t!]{3.9cm}{$h_{zy} + a_{iyzi} - a_{xyxz} - a_{yyyz} - 2a_{zyzz} + c_{ixzz} - c_{izzx} - c_{xixx} - c_{yiyx} - 2c_{zizx}$} & \parbox[t!]{6.5cm}{\begin{align*}
          W_{0}[\rho] = &\{-\frac{i}{2}(\rho_{01}-\rho_{10})X+\frac{1}{2}(\rho_{01}+\rho_{10})Y,\\
           & \frac{i}{2}(\rho_{01}-\rho_{10})X-\frac{1}{2}(\rho_{01}+ \rho_{10})Y \}
          \end{align*}} & $\Delta p_0 = \Delta p_1 =0$\\ 
    \midrule
         $w_1$ & $h_{iy} $ &\parbox[t!]{3.9cm}{$h_{iy} - 2a_{iyzz} - a_{xiyx} - a_{xxyi} - a_{zizy} - c_{ixiz} - 2c_{ixzi} + c_{xyyz} - c_{xzyy} + c_{zxzz}$} & \parbox[t!]{6.5cm}{\begin{align*}
           W_{1}[\rho] = &\{\frac{i}{2}(\rho_{01}-\rho_{10})X-\frac{1}{2}(\rho_{01}+ \rho_{10})Y, \\
           & \frac{i}{2}(\rho_{01}-\rho_{10})X-\frac{1}{2}(\rho_{01}+ \rho_{10})Y\}
          \end{align*}} & $\Delta p_0 = \Delta p_1 =0$\\ 
    \midrule
         $w_2$ & $h_{zx}$ &\parbox[t!]{3.9cm}{$h_{zx} + a_{ixzi} - a_{xxxz} - a_{yxyz} - 2 a_{zxzz} - c_{iyzz} + c_{izzy} + c_{xixy} + c_{yiyy} + 2c_{zizy}$} & \parbox[t!]{6.5cm}{\begin{align*}
           W_{2}[\rho] =&\{-\frac{1}{2} (\rho_{01}+\rho_{10})X-\frac{i}{2}(\rho_{01}-\rho_{10})Y,\\
           & \frac{1}{2}(\rho_{01}+\rho_{10})X+\frac{i}{2}(\rho_{01}-\rho_{10})Y \}
          \end{align*}} & $\Delta p_0 = \Delta p_1 =0$\\ 
    \midrule
         $w_3$ & $h_{ix}$ &\parbox[t!]{3.9cm}{$h_{ix} - 2a_{ixzz} + a_{xiyy} + a_{xyyi} - a_{zizx} + c_{iyiz} + 2c_{iyzi} + c_{xxyz} - c_{xzyx} - c_{zyzz}$} & \parbox[t!]{6.5cm}{\begin{align*}
           W_{3}[\rho] =&\{ \frac{1}{2}(\rho_{01}-\rho_{10})X+\frac{i}{2}(\rho_{01}+\rho_{10})Y, \\
           & \frac{1}{2}(\rho_{01}-\rho_{10})X+\frac{i}{2}(\rho_{01}+\rho_{10})Y\} 
          \end{align*}} & $\Delta p_0 = \Delta p_1 =0$\\ 
    \bottomrule
    \end{tabular}
    \caption{\textbf{Unitary weakness}: weakness generated by a unitary error process acting on the CNOT in the gadget picture (Fig~\ref{fig:auxiliary_pic}). Of these errors, all but $W_0$ also perturb the anti-commuting observables. In each case, we report a representative term in the FOMGI quantity in addition to the label, which highlights how we determine the action of a given elementary deviation.  To illuminate the structure of this decomposition, we show the unit action ($\Lambda[\rho] =\{\Lambda_0[\rho], \Lambda_1[\rho] \}$) and unit change in probabilities ($\Delta p_c = \text{Tr}(\Lambda_c[\rho])$). For a elementary deviation $\Lambda$ acting with strength $\lambda$, the output state is $\lambda\Lambda_c[\rho]$ and the change in measurement probabilities is $\text{Tr}(\lambda\Lambda_c[\rho])$.}
    \label{tab:coherence}
\end{table}
\begin{table}[H]
    \centering
    \begin{tabular}{@{} p{0.8cm} p{0.8cm}p{3.9cm} p{6.7cm}p{3.5cm}@{}l}
    \toprule
     Label ($\lambda$) & Rep. term& FOMGI quantity & Action ($\Lambda = \{\Lambda_0, \Lambda_1 \}$)  & Change in probs. ($\Delta p_c = \text{Tr}(\Lambda_c[\rho])$) \\
    \midrule
        $\tilde{w}_0$& $c_{xxyi}$&\parbox[t!]{3.9cm}{$-a_{xxxz} + a_{xyyz} - a_{xzyy} + a_{yxyz} + c_{xixy} + c_{xiyx} + c_{xxyi} - c_{yiyy }$} & \parbox[t!]{6.5cm}{\begin{align*}
          \tilde{W}_{0}[\rho] =& \{\frac{i}{2}(\rho_{01}-\rho_{10})X+\frac{1}{2}(\rho_{01}+ \rho_{10})Y, \\
           &\frac{i}{2}(\rho_{01}+\rho_{10})X-\frac{1}{2}(\rho_{01}+\rho_{10})Y\}
          \end{align*}} & $\Delta p_0 = \Delta p_1 = 0$\\ 
    \midrule
        $\tilde{w}_1$ &$c_{xiyy} $ &\parbox[t!]{3.9cm}{$-a_{xxyz} - a_{xyxz} + a_{xzyx} + a_{yyyz} - c_{xixx} + c_{xiyy} + c_{xyyi} + c_{yiyx}$} &  \parbox[t!]{6.5cm}{\begin{align*}
         \tilde{W}_{1}[\rho] &= \{-\frac{1}{2}(\rho_{01}+\rho_{10})X+\frac{i}{2}(\rho_{01}- \rho_{10})Y, \\
          & -\frac{1}{2}(\rho_{01}+\rho_{10})X+\frac{i}{2}(\rho_{01}- \rho_{10})Y\}\end{align*}} & $\Delta p_0 = \Delta p_1 = 0$\\  
    \midrule
    $\tilde{w}_2$ & $c_{yyyz}$ &\parbox[t!]{3.9cm}{$-a_{xixx} + a_{xiyy} - a_{xyyi} + a_{yiyx} - c_{xxyz} - c_{xyxz} - c_{xzyx} + c_{yyyz}$} &  \parbox[t!]{6.5cm}{\begin{align*}
          \tilde{W}_{2}[\rho] &= \{-\frac{i}{2}(\rho_{01}-\rho_{10})X-\frac{1}{2}(\rho_{01}+\rho_{10})Y, \\
          &(\frac{i}{2}\rho_{01}-\rho_{10})X+\frac{1}{2}(\rho_{01}+\rho_{10})Y\} 
          \end{align*}} & $\Delta p_0 = \Delta p_1 = 0$\\  \midrule
    $\tilde{w}_3$ &  $c_{xxxz}$ &\parbox[t!]{3.9cm}{$-a_{xixy} - a_{xiyx} + a_{xxyi} + a_{yiyy} + c_{xxxz} - c_{xyyz} - c_{xzyy} - c_{yxyz} $
          } & \parbox[t!]{6.5cm}{\begin{align*}
          \tilde{W}_{3}[\rho] =& \{\frac{1}{2}(\rho_{01}+\rho_{10})X-\frac{i}{2}(\rho_{01}-\rho_{10})Y, \\
          &-\frac{1}{2}(\rho_{01}+\rho_{10})X+\frac{i}{2}(\rho_{01}-\rho_{10})Y\} 
          \end{align*}} & $\Delta p_0 = \Delta p_1 = 0$\\ 
    \bottomrule
    \end{tabular}
    \caption{\textbf{Non-unitary weakness}: weakness generated by a non-unitary error process acting on the CNOT in the gadget picture (Fig~\ref{fig:auxiliary_pic}). In each case, we report a representative term in the FOMGI quantity in addition to the label, which highlights how we determine the action of a given elementary deviation. For a elementary deviation $\Lambda$ acting with strength $\lambda$, the output state is $\lambda\Lambda_c[\rho]$ and the change in measurement probabilities is $\text{Tr}(\lambda\Lambda_c[\rho])$.}
    \label{tab:weak_counterparts}
\end{table}

\section{Extracting FOMGI quantities from a QI}
Below, we explicitly outline how to extract FOMGI quantities from a QI. Suppose we have a noisy QI  $\mathcal{Q}= \bar{\mathcal{Q}} + \Delta \mathcal{Q} $, where:

\begin{equation}
\Delta \mathcal{Q} = \left\{
\begin{pmatrix}
q_0^{II} & q_0^{IX} & q_0^{IY} & q_0^{IZ} \\
q_0^{XI} & q_0^{XX} & q_0^{XY} & q_0^{XZ} \\
q_0^{YI} & q_0^{YX} & q_0^{YY} & q_0^{YZ} \\
q_0^{ZI} & q_0^{ZX} & q_0^{ZY} & q_0^{ZZ}
\end{pmatrix},
\begin{pmatrix}
q_1^{II} & q_1^{IX} & q_1^{IY} & q_1^{IZ} \\
q_1^{XI} & q_1^{XX} & q_1^{XY} & q_1^{XZ} \\
q_1^{YI} & q_1^{YX} & q_1^{YY} & q_1^{YZ} \\
q_1^{ZI} & q_1^{ZX} & q_1^{ZY} & q_1^{ZZ}
\end{pmatrix}
\right\}
\end{equation}
We can construct a vector,

\begin{equation}
    \vec{q} = \begin{pmatrix}
q_0^{II} \\ 
q_0^{IX} \\ 
\vdots \\ 
q_0^{ZZ} \\ 
q_1^{XI} \\
\vdots \\
q_1^{ZZ}
\end{pmatrix}
\end{equation}
from the independent matrix elements (parameters) of $\Delta \mathcal{Q}$. This vector has a length of $2 \times 4^2 - 4 = 28$  since the top row must sum to $\left( 1, 0, 0, 0 \right)$ if the QI is trace-preserving (TP). This means that there do not exist independent $q_1^{II}$, $q_1^{IX}$ and $q_1^{IY}$, and $q_1^{IZ}$ matrix elements. When expressed in this manner, the FOMGI deviations form a basis for this vector space. The error strengths, which correspond to the values of the FOMGI quantities, can be extracted by performing a change of basis on $\vec{q}$. The change-of-basis matrix is 
\begin{equation}
F = \begin{pmatrix}
| & | & \hdots & | \\
\vec{f}_0 & \vec{f}_1 & \hdots & \vec{f}_{27} \\
| & | & \hdots & | \\
\end{pmatrix},
\end{equation}
where $\vec{f}_i$ is the vectorized $i$-th elementary deviation. The error strengths $\vec{s}$ are then given by: 

\begin{equation}
    \vec{s} = F^{-1}\vec{q}. \\ 
\end{equation}
Here, the rows of $F^{-1}$ form the dual basis. Much as is the case with gate error generators \cite{blume-kohout2022a}, not all elementary deviations are mutually orthogonal.\footnote{With a small amount of effort, it is possible to construct a mutually orthogonal basis by swapping some of the primal vectors with their dual counterparts. The practical benefits and tradeoffs of such a mutually orthogonal basis remain to be explored.} The majority of the dual vectors are given by: 

\begin{equation}
    (\Lambda_i)^{\mathrm{D}}[\rho] = \frac{1}{d^2}\Lambda_i[\rho], 
\end{equation}
where $d=2$ is the dimension of the Hilbert space. In these cases, both the dual vectors and primal vectors are mutually orthogonal. There are some deviations for which this is not the case, and the dual vector is not proportional to the primal vector: 

\begin{equation}
\begin{split}
    (R_x^{\mathrm{meas}})^\mathrm{D}[\rho] &=\frac{1}{d^2}( \frac{3}{2}R_x^{\mathrm{meas}}[\rho]- \frac{1}{2}\tilde{R}_x^{\mathrm{meas}}[\rho])\\ 
    (\tilde{R}_x^{\mathrm{meas}})^\mathrm{D}[\rho] &= \frac{1}{d^2}(\frac{3}{2} \tilde{R}_x^{\mathrm{meas}}[\rho]-\frac{1}{2}R_x^{\mathrm{meas}}[\rho])\\ 
    (\tilde{R}_{x,z}^{\mathrm{meas}})^\mathrm{D}[\rho] &= \frac{1}{d^2}(\tilde{R}_{x,z}^{\mathrm{meas}}[\rho]-\frac{1}{2}\tilde{R}_x^{\mathrm{meas}}[\rho]-\frac{1}{2}R^x_{\mathrm{meas}}[\rho])\\
     (R_y^{\mathrm{meas}})^\mathrm{D}[\rho] &=\frac{1}{d^2}( \frac{3}{2}R_y^{\mathrm{meas}}[\rho]- \frac{1}{2}\tilde{R}_y^{\mathrm{meas}}[\rho])\\ 
    (\tilde{R}_y^{\mathrm{meas}})^\mathrm{D}[\rho] &= \frac{1}{d^2}(\frac{3}{2} \tilde{R}_y^{\mathrm{meas}}[\rho]-\frac{1}{2}R_y^{\mathrm{meas}}[\rho])\\
    (\tilde{R}_{y,z}^{\mathrm{meas}})^\mathrm{D}[\rho] &= \frac{1}{d^2}(\tilde{R}_{y,z}^{\mathrm{meas}}[\rho]-\frac{1}{2}\tilde{R}_y^{\mathrm{meas}}[\rho]-\frac{1}{2}R_y^{\mathrm{meas}}[\rho])\\
\end{split}
\end{equation}
\begin{equation}
\begin{split}
    (S^{\mathrm{meas}})^\mathrm{D}[\rho] &= \frac{1}{d^2}(2S_{\mathrm{meas}}[\rho]-S_{\mathrm{prep}}[\rho])\\
    (S^{\mathrm{prep}})^\mathrm{D}[\rho] &= \frac{1}{d^2}(2S^{\mathrm{prep}}[\rho]-S^{\mathrm{meas}}[\rho]-S^{\mathrm{read}}[\rho])\\
    (S^{\mathrm{read}})^\mathrm{D}[\rho] &= \frac{1}{d^2}(2S^{\mathrm{read}}[\rho]-S^{\mathrm{prep}}[\rho]) \\
    \end{split}
\end{equation}
\begin{equation}
\begin{split}
    (A^{\mathrm{meas}})^\mathrm{D}[\rho] &= \frac{1}{d^2}(A^{\mathrm{meas}}[\rho]-\frac{1}{2}A^{\mathrm{prep}}[\rho])\\
    (A^{\mathrm{prep}})^\mathrm{D}[\rho] &= \frac{1}{d^2}(A^{\mathrm{prep}}[\rho]-\frac{1}{2}A^{\mathrm{meas}}[\rho]-\frac{1}{2}A^{\mathrm{read}}[\rho])\\
    (A^{\mathrm{read}})^\mathrm{D}[\rho] &= \frac{1}{d^2}(A^{\mathrm{read}}[\rho]-\frac{1}{2}A^{\mathrm{prep}}[\rho]) \\
\end{split}
\end{equation}

\section{Leakage quantification and post-selection}\label{appendix:leakage_postselection}

\begin{figure}
    \centering
    \includegraphics[width=1.0\linewidth]{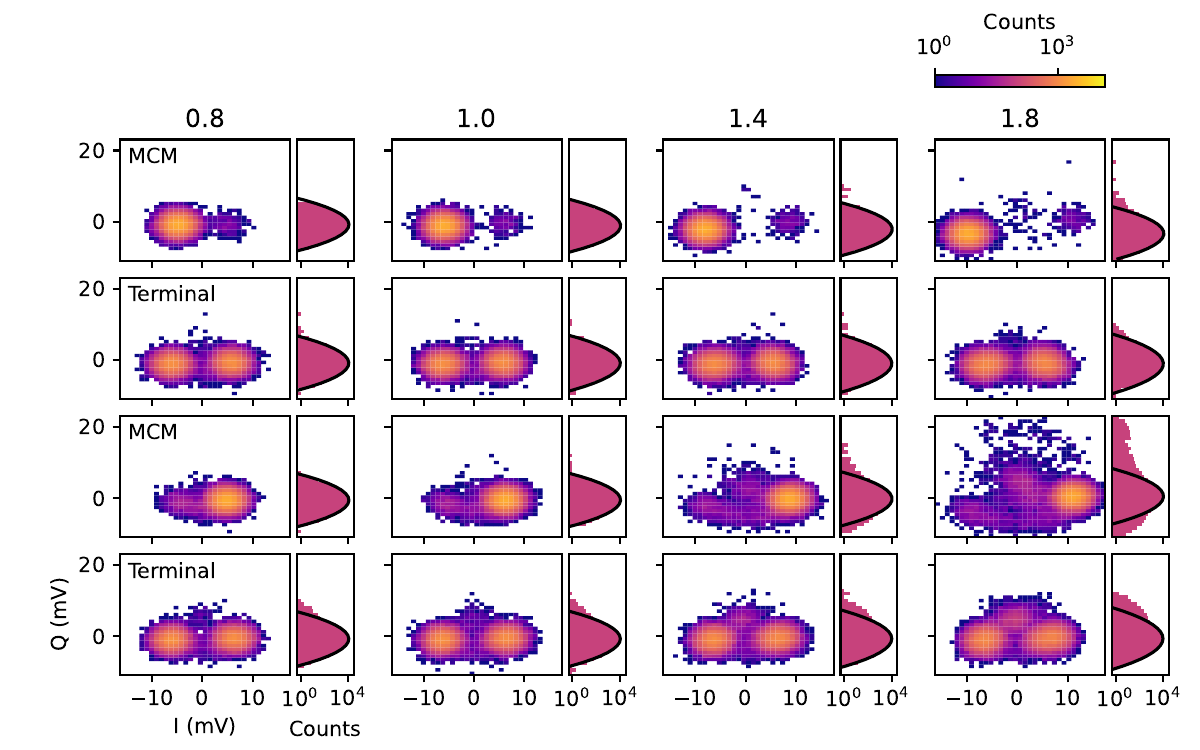}
    \caption{
    \textbf{Leakage in raw GST data.}
    Histograms of IQ voltages for a subset of GST circuits at different MCM amplitudes $V$.
    Column labels indicate $ V/ V_0$ (e.g., the second column is the nominal readout setting). \textbf{First row}: Histogram of MCM outcomes combined for the six GST circuits with an expected definite MCM outcome of $|0\rangle$. \textbf{Second row: }Histogram of terminal measurement outcomes for the same circuits as in the first row.
    \textbf{Third row:} MCM outcomes for the six GST circuits with an expected definite MCM outcome of $|1\rangle$. \textbf{Fourth row:} Terminal measurement outcomes for the same circuits as in the third row. One-dimensional histograms in pink are projected onto the Q axis, which has minimal discriminating power between $|0\rangle$ and $|1\rangle$. Terminal outcome distributions are approximately equal in all cases, with half $|0\rangle$ and half $|1\rangle$ because the six circuits average over a complete set of measurement fiducials. Each sub-figure includes $6 \times 8000 = 32000$ shots.}
    \label{fig:mcm_leakage}
\end{figure}

In this appendix, we examine the raw IQ data at $V\ge 0.8V_0$ and describe a procedure for removing shots where leakage was likely to have occurred. The cluster in Fig.~\ref{fig:error_mechanisms}(b) that emerges at larger amplitude $V$ is consistent with the location expected for shifts of the readout resonator greater than $\chi_{01}$. Therefore, we believe it corresponds to leakage out of the computational subspace. Given this device's parameters, it is not possible to ascertain from this data whether leakage events are to the transmon $|2\rangle$ state, or to some higher excited state of the transmon. For the purposes of this analysis, we can consider all such states to be a single incoherent ``Leaked'' manifold.

\begin{figure}
    \centering
    \includegraphics[width=1.0\linewidth]{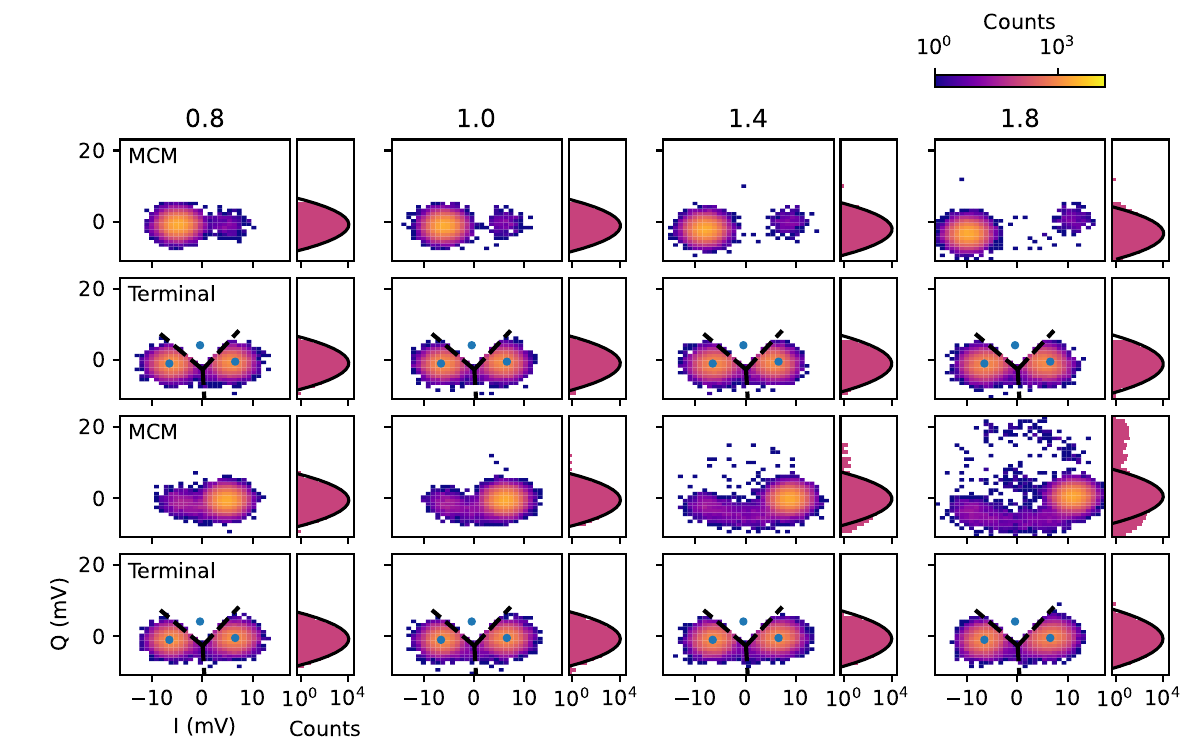}
    \caption{\textbf{Leakage post-selection.}
    Same data as Fig.~\ref{fig:mcm_leakage}, with shots classified as ``Leaked'' in the terminal measurement removed from both MCM and terminal plots.
    Circles show the minimum-distance classifier centroids used for the terminal measurement; black dashed lines are the classifier boundaries.}
    \label{fig:mcm_leakage_postselection}
\end{figure}

Figure~\ref{fig:mcm_leakage} shows leakage events in the raw data from a subset of our GST experiments.
Since leakage is observed to be preferentially out of the $|1\rangle$ state, we present data from all circuits that ideally produce a 0 outcome for the MCM (``0 circuits'') and compare them to those that ideally produce 1 (``1 circuits'') for a few different values of measurement amplitude $V$. We observe a strong qualitative difference between the 0 circuits and 1 circuits. In the 0 circuits, very few leakage counts are observed in either the MCM or terminal measurement outcomes at any amplitude, with a very slight increase at the highest amplitudes (note the logarithmic scale). However, the 1 circuits exhibit leakage effects in both measurements that increase strongly with increasing $V$. The small amount of leakage observed in the terminal outcomes even at small readout amplitudes may be due to thermal excitations from $|1\rangle$ to $|2\rangle$.

Due to the long lifetime of the leakage---specifically, the low probability of returning to the computational subspace (seepage) over the timescales of our circuits is small---leakage events in the MCM should be correlated with measured leakage observed in the terminal measurement. To demonstrate this, we implement the post-selection technique used in Section~\ref{ssec:leakage} of the main text. For the terminal measurement dataset with the most pronounced leakage ($V = 1.8\times V_0,$ last sub-figure in Fig.~\ref{fig:mcm_leakage}), we train a three-outcome minimum distance classifier using k-means clustering. We can then use this classifier on all terminal measurements to remove likely leakage events. In Fig.~\ref{fig:mcm_leakage_postselection}, we show the same datasets as Fig.~\ref{fig:mcm_leakage} but with all counts assigned ``Leaked'' in the terminal measurements removed. There is a sharp reduction in the leakage counts in the MCM, demonstrating strong correlation between leakage outcomes in the MCM and terminal measurements.

\begin{figure}
    \centering
    \includegraphics[width=0.6\linewidth]{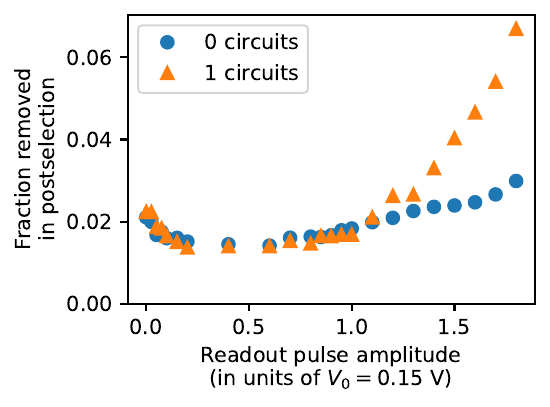}
    \caption{\textbf{Leakage post-selection rate.}
    Fraction of total shots removed in the post-selection process as a function of readout pulse amplitude.
    }
    \label{fig:postselection_rate}
\end{figure}

The post-selection removal rate is plotted in Fig.~\ref{fig:postselection_rate}.
There is a strong divergence between the post-selection rate for 0 circuits and 1 circuits at $ V > V_0$.
The slight increase in post-selection rate at the lowest readout amplitudes is likely due to small drifts in the overall readout phase for the terminal measurement over the 12 hour runtime of the full suite of experiments, since the classifier used is trained on data from the final dataset at $ V = 1.8\times V_0$.

\section{Details of fitting an AC Stark shift model}\label{appendix:stark}
\begin{figure}
    \centering
    \includegraphics[width=0.6\linewidth]{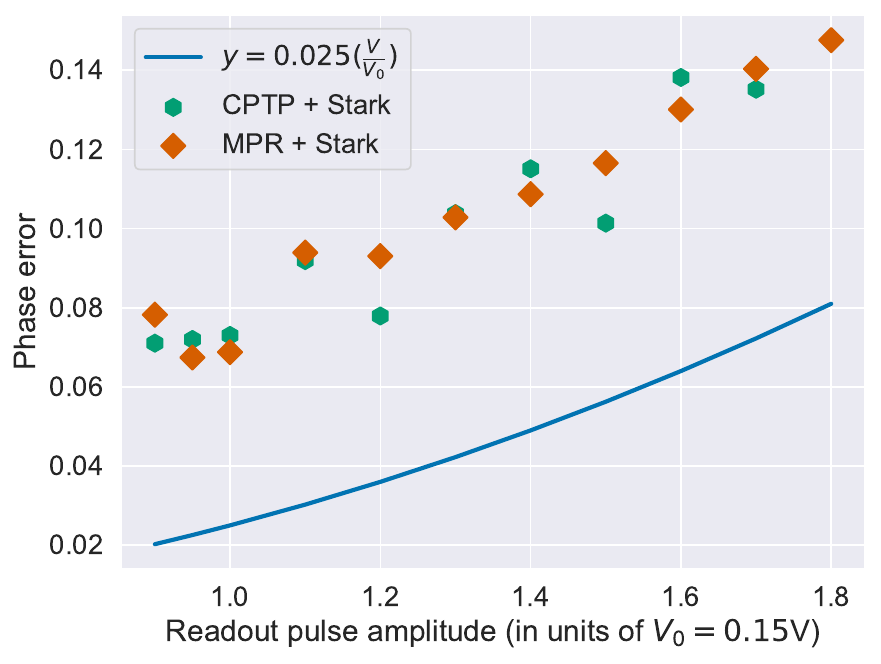}
    \caption{\textbf{Absolute value of the additional phase error $\phi$ as a function of amplitude scale factor $V$. }For both Stark shift models, we observe the correct quadratic scaling, but roughly a factor of two difference in numerical values.}
    \label{fig:stark_shift_estimates}
\end{figure}

The phase error induced by the AC Stark shift can be described according to the following equation \cite{rudinger2022a}, where $n_i$ is the photon number in a given qubit state $| i\rangle$ after the 2 $\upmu$s post-measurement delay, $\kappa$ is the energy decay rate of the readout resonator, $\chi_\mathrm{01}$ is the dispersive shift, and $t_{\mathrm{gate}}$ is the duration of the gates $G_X$ and $G_Y$.  

\begin{equation}
    \phi_{i,m} = n_i \frac{\chi_\mathrm{01}}{2\kappa}\left(1-e^{-\kappa t_{\mathrm{gate}}}\right)e^{\kappa mt_{\mathrm{gate}}}
\end{equation}
The gate time is much smaller than the energy $ t_{\mathrm{gate}} << 1/\kappa.$ This allows us to approximate the phase error as constant on all post-MCM gates and drop the $e^{\kappa mt_{\mathrm{gate}}}$ term, and to take the first-order expansion of the $(1-e^{-\kappa t_{\mathrm{gate}}})$ term.
Measurements of the cavity photon population \added{using the Ramsey technique of} \cite{McClure2016rapid} indicated that the number of photons was independent of outcome (i.e., $n_0 \approx n_1$),
so we replace $n_i$ with $n$, the outcome-averaged number of photons, to arrive at 

\begin{equation}
    \phi \approx n \frac{\chi_\mathrm{01} t_{\mathrm{gate}}}{2}.
\end{equation}
These assumptions justify the use of only a single additional parameter in the Stark shift models, which we choose to model as a post-gate phase error. For the nominal amplitude setting $V = V_0$, we measure $n \approx 1$. As we discuss in Section~\ref{sec:goodness_of_fit}, the occupation of the readout resonator scales as $V^2$. The product of the dispersive shift and the gate time is $\chi_\mathrm{01} t_{\mathrm{gate}} \approx 0.05$, so we predict
\begin{equation}
    \phi \approx 0.025 \left( \frac{V}{ V_0} \right)^2.
\end{equation}
Figure~\ref{fig:stark_shift_estimates} compares the theoretical prediction with the GST estimates.
The correct scaling is obtained, but an overall numerical scale of roughly two is observed, the reason for which is not fully understood. 
This may be due in part to the presence of the additional error mechanism of dephasing due to residual photons, which scales the same way with $V$ \cite{Gambetta2006qubit}.
The coherent phase error parameter may capture both effects.

\section{Experimental parameters}\label{experimental_details}
The experiments were conducted using a chip with 6 uncoupled fixed-frequency transmons anchored to the base plate of a dilution refrigerator at 15 mK. We choose the qubit with the best coherence properties for this study. Each qubit has a dedicated readout resonator in a hanger geometry on a common transmission line. Qubit control pulses are delivered through the common transmission line after being combined with the readout drive at room temperature. Single qubit gates are implemented using 30-ns duration raised cosine shaped pulses with DRAG correction. Readout signals are amplified by a nearly-quantum limited Josephson traveling wave parametric amplifier (TWPA) at 15 mK before being further amplified at 4 K and room temperature before acquisition. Readout signals are acquired, demodulated, and integrated for 1.98 us with constant weighting in time. The system Hamiltonian is defined as 
\begin{equation}
    H = \frac{\omega_q}{2} \sigma_z  + \frac{\omega_0  + \omega_1}{2}a^\dagger a  + \frac{\chi_\mathrm{01}}{2} a^\dagger a \sigma_z 
\end{equation}
and the measured parameters are summarized in Table~\ref{tab:device_parameters}.

\begin{table}[h!]
    \centering

    \begin{tabular}{p{6.5cm} p{2.5cm} p{4cm}}
    \toprule
     Device parameter & Value & Notes \\ 
    \midrule
         Qubit transition frequency ($\omega_\mathrm{q} / 2\pi$) & 3316 MHz & \\ 
         Resonator frequency ($\omega_0 / 2\pi$) & 7139 MHz & As measured with qubit in ground state \\ 
         Qubit excited state lifetime ($T_1$) & 90 $\upmu$s  & \\ 
         Resonator energy decay rate ($\kappa / 2\pi$) & 0.160 MHz  & \\ 
         Dispersive shift ($\chi_\mathrm{01}/ 2\pi$) & $-$0.260 MHz & Full dispersive shift, e.g.  $ \chi_\mathrm{01} = \omega_1  - \omega_0 $\\ 
    \bottomrule
    \end{tabular}
    \caption{Transmon qubit device parameters.}
    \label{tab:device_parameters}
\end{table}

\section{Estimated FOMGI quantities versus readout amplitude}
\begin{figure}
    \centering
    \includegraphics[width=0.999\linewidth]{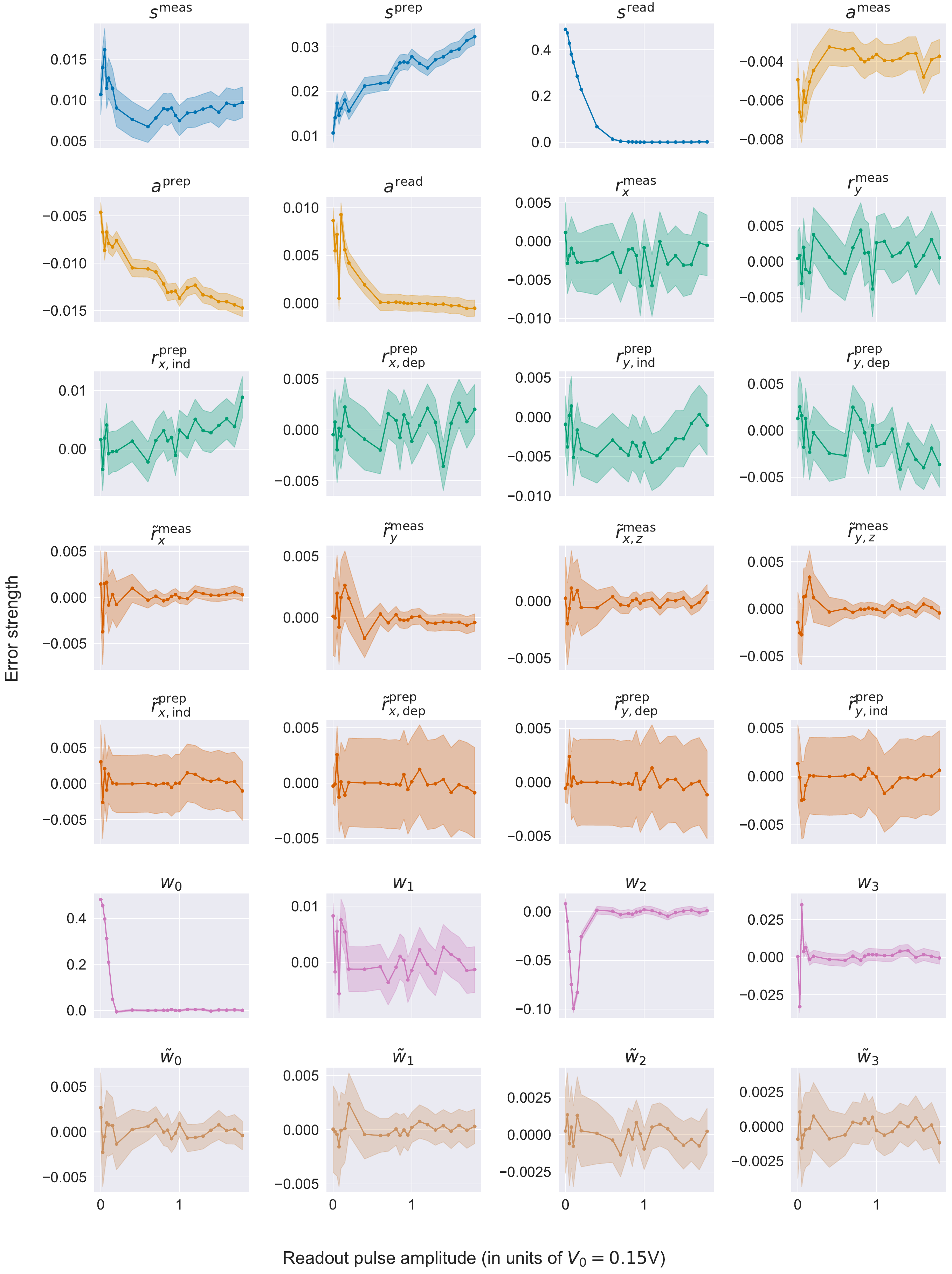}
    \caption{\textbf{Extracted FOMGI quantities as a function of $V$.} Colors indicate the classes of MCM errors. In order, these are Pauli-stochastic (blue), active (yellow), axis rotation (green), $\rho$-dependent axis rotation (orange), unitary weakness (pink), and non-unitary weakness (bronze).}
    \label{fig:FOMGI_as_function_of_alpha}
\end{figure}

In Fig.~\ref{fig:FOMGI_as_function_of_alpha}, we show the all estimated FOMGI quantities as a function of $V$ for the CPTP model. The most experimentally significant of these FOMGI quantities (often shown as sums and differences as detailed in Appendix~\ref{appendix:classification}) are discussed in Section~\ref{sec:error_decomp}. 

\section{Full GST results for high-precision dataset}\label{app:gst_results}

We include full GST estimates of the gate set for the $V=V_0$ and $N=4\times 10^{4}$ dataset in Table~\ref{tab:GST_results}, including the MCM, for the nominal amplitude $V = V_0$ experiment with $N=4 \times 10^{4}$ shots. These estimates were fit using the measure-and-prepare model that was augmented to capture pure readout error (``MPR''), as described in Section~\ref{sec:goodness_of_fit}. Consequently, the inner sub-block of matrix elements---capturing weakness at first order---round to zero.  As expected, we also observe that the single-qubit gate errors are significantly smaller than the MCM errors.
\begin{table}[ht]
    \centering
    \begin{tabular}{@{}p{1.5cm} p{3.4cm} p{6cm} p{5cm}@{}}
    \toprule
    Operation & Target & Estimate & $2\sigma$ error bars ($+/-$)\\ 
    \midrule
         $|\rho\rangle\!\rangle$ & 
         \[\begin{pmatrix}
             \frac{1}{\sqrt{2}} \\  0 \\ 0 \\ \frac{1}{\sqrt{2}}
       \end{pmatrix}\] &  
         \[\begin{pmatrix}
             0.707 \\ 0.004 \\ 0.003 \\ 0.702
         \end{pmatrix}\] & \[\begin{pmatrix}
             0.0 \\ 0.002 \\ 0.002 \\ 0.0
         \end{pmatrix}\] \\ 
    \midrule
         $\langle\!\langle\hat{M}|$ & \[\begin{pmatrix}
             \frac{1}{\sqrt{2}} &  0 & 0 & \frac{1}{\sqrt{2}}
       \end{pmatrix}\] & \[\begin{pmatrix}
             0.719 &  0.001 & 0.002 & 0.695
       \end{pmatrix}\] & \[\begin{pmatrix}
             0.0 &  0.002 & 0.002 & 0.0
       \end{pmatrix}\] \\
    \midrule
         $\hat{G}_X$ & 
         \[
\begin{pmatrix}
1 & 0 & 0 & 0 \\
0 & 1 & 0 & 0 \\
0 & 0 & 0 & -1.0 \\
0 & 0 & 1.0 & 0 \\
\end{pmatrix}
\] & 
\[
\begin{pmatrix}
1.0 & 0.0 & 0.0 & 0.0 \\
0.0 & 0.999 & 0.001 & 0.003 \\
0.0 & 0.002 & -0.003 & -0.999 \\
0.0 & 0.0 & 0.999 & -0.003
\end{pmatrix}
\]  & 
\[
\begin{pmatrix}
0.0 & 0.0 & 0.0 & 0.0 \\
0.001 & 0.001 & 0.003 & 0.003 \\
0.001 & 0.003 & 0.002 & 0.0 \\
0.001 & 0.003 & 0.0 & 0.001
\end{pmatrix}
\] \\ 
    \midrule
         $\hat{G}_Y$ & 
         \[
\begin{pmatrix}
1.0 & 0 & 0 & 0 \\
0 & 0 & 0 & 1 \\
0 & 0 & 1 & 0 \\
0 & -1 & 0 & 0 \\
\end{pmatrix}
\] & 
\[
\begin{pmatrix}
1.0 & 0.0 & 0.0 & 0.0 \\
0.0 & -0.003 & 0.0 & 0.999 \\
0.0 & 0.003 & 1.0 & 0.0 \\
-0.001 & -0.999 & 0.004 & -0.003
\end{pmatrix}
\] &
\[
\begin{pmatrix}
0.0 & 0.0 & 0.0 & 0.0 \\
0.001 & 0.003 & 0.003 & 0.0 \\
0.001 & 0.003 & 0.001 & 0.003 \\
0.001 & 0.0 & 0.003 & 0.001
\end{pmatrix}
\] \\ 

\midrule
         $\hat{G}_I$ & 
         \[
\begin{pmatrix}
1 & 0 & 0 & 0 \\
0 & 1 & 0 & 0 \\
0 & 0 & 1 & 0 \\
0 & 0 & 0 & 1 \\
\end{pmatrix}
\] & 
\[
\begin{pmatrix}
1.0 & 0.0 & 0.0 & 0.0 \\
-0.001 & 0.946 & -0.037 & -0.004 \\
0.002 & 0.046 & 0.947 & 0.0 \\
0.073 & 0.009 & -0.005 & 0.926
\end{pmatrix}
\] &
\[
\begin{pmatrix}
0.0 & 0.0 & 0.0 & 0.0 \\
0.002 & 0.002 & 0.006 & 0.006 \\
0.002 & 0.006 & 0.002 & 0.006 \\
0.002 & 0.006 & 0.006 & 0.002
\end{pmatrix}
\] \\ 
    \midrule
         $\hat{\mathcal{Q}}_0$& 
         \[
\begin{pmatrix}
0.5 & 0 & 0 & 0.5 \\
0 & 0 & 0 & 0 \\
0 & 0 & 0 & 0 \\
0.5 & 0 & 0 & 0.5 \\
\end{pmatrix}
\] & 
\[
\begin{pmatrix}
0.507 & 0.004 & 0.001 & 0.492 \\
-0.003 & 0.0 & 0.0 & -0.003 \\
0.003 & 0.0 & 0.0 & 0.003 \\
0.506 & 0.004 & 0.001 & 0.491
\end{pmatrix}
\]  & \[
\begin{pmatrix}
0.0 & 0.001 & 0.001 & 0.0 \\
0.002 & 0.004 & 0.004 & 0.002 \\
0.002 & 0.004 & 0.004 & 0.002 \\
0.0 & 0.001 & 0.002 & 0.001
\end{pmatrix}
\] \\ 
    \midrule
         $\hat{\mathcal{Q}}_1$& 
         \[
\begin{pmatrix}
0.5 & 0 & 0 & -0.5 \\
0 & 0 & 0 & 0 \\
0 & 0 & 0 & 0 \\
-0.5 & 0 & 0 & 0.5 \\
\end{pmatrix}
\] & 
\[
\begin{pmatrix}
0.493 & -0.004 & -0.001 & -0.492 \\
0.0 & 0.0 & 0.0 & 0.0 \\
-0.002 & 0.0 & 0.0 & 0.002 \\
-0.434 & 0.003 & 0.001 & 0.434
\end{pmatrix}
\] & \[
\begin{pmatrix}
0.0 & 0.001 & 0.001 & 0.0 \\
0.001 & 0.004 & 0.004 & 0.001 \\
0.001 & 0.004 & 0.004 & 0.001 \\
0.001 & 0.002 & 0.002 & 0.001
\end{pmatrix}
\] \\ 
    \bottomrule
    \end{tabular}
    \caption{GST estimates for all operations for dataset with $V = V_0$ and $N=4 \times 10^4$ shots.}
    \label{tab:GST_results}
\end{table}

\end{document}